\documentclass[twocolumn,hyperpdf,amsmath,amssymb,aps,prd,10pt,superscriptaddress,nofootinbib,noeprint,preprintnumbers,floatfix]{revtex4-2}
\pdfoutput=1

\usepackage[utf8]{inputenc} 

\usepackage{graphicx, color}

\usepackage[letterspace=-10]{microtype} 

\usepackage{bm, amsmath, amsfonts}
\usepackage{mathtools}
\usepackage{algorithm,algorithmic}
\usepackage{braket}
\usepackage{amssymb}
\usepackage{multirow, tabularx, dcolumn}
\usepackage{mathtools} 
\usepackage{booktabs}
\usepackage{placeins}
\usepackage{subcaption}
\usepackage{epstopdf}
\usepackage{comment}
\usepackage[colorinlistoftodos]{todonotes}

\usepackage{hyperref}

\usepackage{xcolor}
\definecolor{jlab_red}{RGB}{192,39,45}
\definecolor{jlab_orange}{RGB}{249,102,0}
\definecolor{jlab_blue}{RGB}{47,122,121}
\definecolor{jlab_green}{RGB}{65,125,10}

\newcommand{\nD}{\ensuremath{n_{\cal D}}}
\newcommand{\vD}{\ensuremath{\Omega_{\cal D}}}
\newcommand{\flowt}{\ensuremath{s}}
\newcommand{\perm}[1]{\ensuremath{P^{\phantom{\dagger}}_{#1}}}
\newcommand{\permDag}[1]{\ensuremath{P^{\dagger}_{#1}}}

\hypersetup{%
pdftitle = {Optimising stochastic algorithms for hadron correlation function 
	computations in lattice QCD using a localised distillation basis},
pdfsubject = {QCD},
pdfkeywords = {QCD, Hadron, Physics, Lattice, Scattering, Distillation},
pdfauthor = {Nicolas Lang, Robert G. Edwards, Michael J. Peardon},
colorlinks = {true},
filecolor = {black},
linkcolor = {jlab_blue},
menucolor = {black},
citecolor = {jlab_green},
urlcolor = {jlab_green},
}{}

\DeclareMathOperator{\EX}{\mathbb{E}}
\DeclareMathOperator{\VAR}{\mathrm{Var}}

\newcommand{\pluseq}{\mathrel{+}=}
\newcommand{\vp}{\ensuremath{\underline{\smash{p}}}}
\newcommand{\vx}{\ensuremath{\underline{\smash{x}}}}
\newcommand{\vy}{\ensuremath{\underline{\smash{y}}}}

\newcommand{\vs}{\ensuremath{\underline{\smash{s}}}}

\begin{document}

\title{Optimising stochastic algorithms for hadron correlation function 
  computations in lattice QCD using a localised distillation basis}

\author{Nicolas Lang}
\email{nicolas.lang@ific.uv.es}
\affiliation{Instituto de F\'{i}sica Corpuscular (IFIC), Universitat de Valencia, 46071, Valencia, Spain}
\author{Robert G. Edwards}\email{edwards@jlab.org}
\affiliation{Thomas Jefferson National Accelerator Facility, 12000 Jefferson Avenue, Newport News, VA 23606, USA}
\author{Michael J. Peardon}\email{mjp@maths.tcd.ie}
\affiliation{School of Mathematics, Trinity College, Dublin 2, Ireland}


\collaboration{for the Hadron Spectrum Collaboration}
\date{\today}

\begin{abstract}
Distillation is a quark-smearing method for the construction of a broad class of 
hadron operators useful in lattice QCD computations and defined via a 
projection operator into a vector space of smooth gauge-covariant fields. A 
new orthonormal basis for this space is constructed which builds in locality. 
This basis is useful for the construction of stochastic methods to estimate the 
correlation functions computed in Monte Carlo calculations relevant for 
hadronic physics. 
\end{abstract}

\maketitle

\noindent

\section{Introduction}

Distillation \cite{Peardon:2009gh} is a useful framework for 
computing correlation functions between a wide range of hadron-interpolating 
operators in lattice QCD. It is a quark smearing acting on the 
fields on a time-slice of the lattice by applying the projection operator into 
\vD, the space spanned by the modes of the three-dimensional gauge-covariant Laplace 
operator on that slice. This technique has proved successful for meson 
spectroscopy (see for example Refs.~
\cite{Dudek:2009qf,Dudek:2011tt,HadronSpectrum:2012gic,Lang:2022elg,Wilson:2023hzu,Yan:2024gwp,Li:2024pfg,Boyle:2024hvv})
where distilled operators have good overlap 
onto the low-lying spectrum of the hamiltonian while enabling a broad range 
of creation operators resembling multi-hadron states to be employed. 
Good overlap \cite{Knechtli:2022bji} onto these states is important in calculations of scattering 
and hadron decays in Euclidean field theory calculations via the L\"uscher
method \cite{Luscher:1990ux,Luscher:1990ck} 
(see Ref.~\cite{Briceno:2017max} for a review). 

The disadvantage of the method is seen when interpolating operators with more 
constituent quark fields such as for baryons 
\cite{Basak:2005aq,Edwards:2011jj,Green:2021qol} or tetraquarks \cite{Cheung:2017tnt,Collins:2024sfi} 
are used. 
Evaluating the diagrams following from Wick contractions of the quarks requires 
$\mathcal{O} (\nD^{d+1})$ arithmetic operations, where \nD\ is the rank of 
distillation space and $d$ the number of constituent fermion fields in the 
operator or equivalently $\mathcal{O} (V_3^{d+1})$, assuming the number of 
distillation vectors is proportional to the spatial volume $V_3$.
One established approach to reduce computational costs is to sample with 
random vectors in \vD\ and construct unbiased estimators for correlation 
functions \cite{Morningstar:2011ka}.  The statistical noise added by this
sampling is further reduced using dilution schemes \cite{Foley:2005ac} and 
effectiveness of the combination has been demonstrated in a range of 
calculations. These remain expensive so it is worth pursuing 
alternatives. 

In this paper, a new basis for distillation space is constructed as a 
potential starting point in the design of stochastic schemes which reduce 
variance. The basis 
vectors are localised near defined positions on a coarse spatial grid and so the
corresponding local hadron operators exhibit an approximate sparsity in their 
tensor distillation-space structure; the components where all constituents are 
near the same basis grid point are substantially larger than others. This 
basis and its numerical evaluation is described in Sec.~II. The sparsity can 
be exploited in stochastic representations of the correlation function and 
first tests of these cost-reductions are described in Secs.~III and IV. Secs. 
V and VI discuss our findings and summarise this work.

\section{A localised basis for distillation space \label{sec:basis}}
Distillation is a linear operator acting on the fermion fields on a 
single time-slice of the lattice which projects into the small vector space of 
smooth modes spanned by the gauge-covariant three-dimensional Laplace 
operator. The distillation operator can be written
\begin{align}
  \Box(t) = V(t) V^\dagger(t),
  \label{eq:Box}
\end{align}
where $V(t)$ is the matrix whose \nD\ columns are the normalised eigenvectors 
with smallest eigenvalues of the 
Laplace operator on time-slice $t$. 
The corresponding \nD-dimensional vector space of smooth modes 
is labelled \vD. 
The operator of Eqn.~\ref{eq:Box}
is invariant under any $U(\nD)$ global unitary transformation of the basis 
vectors, $V'=V U$ with $U U^\dagger=I$. 

The two-point correlation function for isovector meson 
creation operator $\Phi$ formed from distilled fields on time-slices $t_0$ and $t_1$ 
is written
\begin{align}
  C(t_1,t_0) = \mbox{Tr } \Phi(t_1) \tau(t_1,t_0) \Phi(t_0) \tau(t_1,t_0), 
   \label{eq:DistCorr}
\end{align}
after defining the $\nD\times\nD$ matrices
\begin{align}
  \Phi(t)       &= V^\dagger(t) \Gamma(t) V(t) \nonumber\\
  \tau(t_1,t_0) &= V^\dagger(t_1) D^{-1}(t_1,t_0) V(t_0) . 
\end{align}
$ \Phi(t)$ encodes the spatial and Dirac structure of the operator and is discussed further below.
$\tau$ is commonly referred to as the \textit{perambulator} and describes 
quark propagation between the distillation spaces on the two time-slices. 
It is clear from Eqn.~\ref{eq:DistCorr} the computation of the meson two-point 
function involves summing over
a large number of contributions from smooth modes on the source and sink time-slices. 

The smearing operator acts locally, spreading the field out at the confining 
scale of hadrons and so can be regarded as a low-rank and approximately 
sparse matrix. The sparsity is not seen in the basis of
eigenvectors of the Laplace operator, which cover the entire 
three-dimensional time-slice. Methods to reduce the computational cost of 
evaluating correlation functions by exploiting locality can be constructed by 
working in a basis where this locality is more apparent. 

The distillation operator is a low-rank projection into the smooth space that is invariant under spatial translations,
so its application to \nD\ point sources evenly distributed on the time-slice 
forms a spanning set for \vD.
This spanning set is localised 
by construction. The locations of the source points, where three independent
coloured degrees of freedom are supported is constrained such that the number 
of sources matches \nD. For example, defining a mesh of $4^3$ source points
evenly spread over the time slice would span a \nD=$3\times 4^3=192$ distillation space. 
The vectors at source point $x$ can be constructed to be covariant under gauge 
transformations by simply copying only the three entries at site $x$ from the 
lowest three Laplace eigenvectors into the sources.
Define $Q$ as the matrix with this grid of point vectors as its columns. 
Now applying the distillation operator yields column matrix $\bar{W}$, a new 
set of $n_D$ vectors which span \vD;
\begin{align}
  \bar{W} = \Box Q = V A_0 \mbox{ with } A_0 = V^\dagger \bar{W} = V^\dagger Q. 
\end{align}
Here, the square matrix $A_0$ represents the mapping from the orthonormal 
eigenvector basis $V$ to the new localised spanning set $\bar{W}$.  Since the 
columns in $\bar{W}$ do
not necessarily form an orthonormal basis, 
$A_0$ is not always unitary. 

Developing software and algorithms is greatly 
simplified if an orthonormal basis for this spanning set can be used, 
but locality of the point source construction needs to be preserved. 
Traditionally, the Gram-Schmidt method is used, where 
basis vector $e_k$ is explicitly orthogonalised against the preceding 
elements ${e_j}, j=1\dots k-1$ by subtraction. However this construction is 
not permutation invariant and destroys the locality added into the spanning set at the start of the process. An alternative method introduces a fictitious 
flow-time co-ordinate, \flowt\ along which $A(\flowt)$ evolves and by setting 
$A_0=A(0)$ to be the initial condition of a flow equation whose fixed points 
are the desired unitary matrices in distillation space, so
\begin{align}
   \lim_{\flowt\rightarrow\infty} A(\flowt)^\dagger A(\flowt) = I_{\cal D}. 
\end{align}
Unitary matrices are the set of zeroes of a non-negative action, $S$ acting 
on complex $\nD\times\nD$ matrices and given by 
\begin{align}
  S(A) = \frac 12 \mbox{tr } \Delta^2(A) \mbox{ with } 
    \Delta(A) = I-A A^\dagger. 
   \label{eq:FlowAction}
\end{align}
A flow can be defined by the first-order differential equation in \flowt,
\begin{align}
   \frac{dA}{d\flowt} = -\partial_\ast S \;\;\mbox{ with } \;\;
        [\partial_\ast S]_{ij} = \frac{\partial S}{\partial[A^\ast]_{ij}}
          = \frac{\partial S}{\partial[A^\dagger]_{ji}}. 
\end{align}
This flow follows the steepest-descent path for the action $S$. Taking the 
definition in Eqn.~\ref{eq:FlowAction}, the 
resulting first-order differential equation becomes 
\begin{align}
   \frac{dA}{d\flowt} = \big(I-A(\flowt)A^\dagger(\flowt)\big) \; A(\flowt),
\end{align}
and since for unitary matrices the right-hand side vanishes, they are fixed 
points of this flow. 
For the typical distillation-space dimensions used in spectroscopy 
calculations, evolving these flow 
equations to their fixed point to find an orthonormal basis is not 
computationally costly. Machine precision is easily achieved using the 
fourth-order Runge-Kutta scheme 
accelerated with an adaptive step-size. 

The advantage of this flow-based scheme over Gram-Schmidt is
seen by considering two initial conditions for the flow $A_0$ and $A'_0$, 
related by 
\begin{align}
   A'_0 = \perm{L} A_0 \perm{R}, 
    \label{eq:permA}
\end{align}
with $\perm{L} \permDag{L}=I_{\cal D}$ and 
$\perm{R} \permDag{R}=I_{\cal D}$ 
defining a pair of arbitrary unitary transformations. Simple examples for 
\perm{L,R} 
would be given by a permutation of the source points used to form the local 
set of vectors spanning \vD. The flow equation and Eqn.~\ref{eq:permA} imply
$A'(s) = \perm{L} A(s) \perm{R}, $
for all \flowt\ so the resulting orthonormal basis vectors inherit the 
same transformation property linking $A_0$ and $A'_0$, including permutations.
This is seen empirically to maintain locality for the orthonormal basis 
vectors generated in contrast to using Gram-Schmidt. 

The new orthonormal basis set is found by setting
\begin{align}
  W = V U \mbox{ with } U = \lim_{s\rightarrow\infty} A(s)
\end{align}
after evolving the flow equation to its fixed point. Now the distillation 
smearing operator can be expressed equivalently as either $\Box=V V^\dagger$ or 
  $\Box = W W^\dagger$. 

\begin{figure*}[!thb]
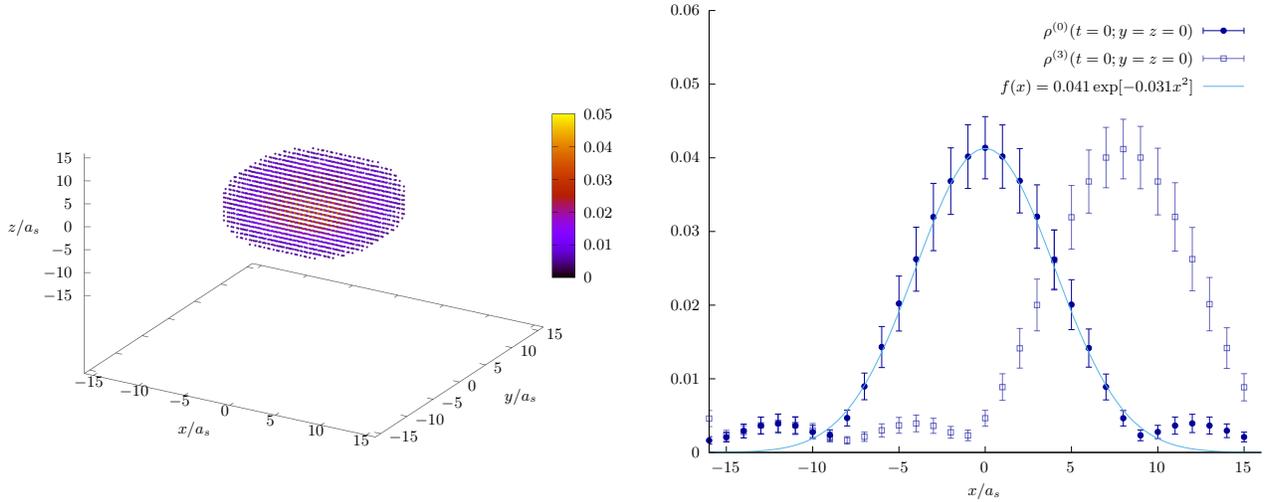

	\begin{subfigure}[l]{0.48\textwidth}
		\resizebox{\textwidth}{!}{
				\input{avg_eig_c0_t0_shifted.tex}
		}
		\label{subfig:dist_lapl_spat}		
	\end{subfigure}
	\begin{subfigure}[l]{0.48\textwidth}
		\resizebox{\textwidth}{!}{
			\input{avg_eig_c0_t0.x_slice.fit+w3.tex}
		}
		\label{subfig:dist_eig_rad}		
	\end{subfigure}
	\caption{Left: spatial distribution of $\rho^{(0)}(\vx,t=0)$. Elements which are smaller than one-eighth of the maximum are suppressed in the plot for improved readability. Right: Slice along the x-axis of the spatial distribution of $\rho^{(i)}(\vx,t=0)$ for $i = 0$ and $i = 3$. For $i=0$ a Gaussian fit was performed and the blue line represents the fit result.
		The distributions correspond to the ensemble average over the magnitudes.}
\label{fig:dist_vectors}
\end{figure*}

The time-slice color field $w^{(i)}_a(\vx,t)$ is formed from the $i$-th column 
of $W$ and 
all \nD\ of these fields are seen to be localized close to their source point 
on time-slice $t$ after computing the local norm, 
\begin{align}
  \rho^{(i)}(\vx,t) = \bigg( \sum_{a=1}^3 
     w^{(i)*}_a(\vx,t)\;
     w^{(i) }_a(\vx,t) \bigg)^{\frac 12}. 
\end{align}
Figure \ref{fig:dist_vectors} shows the gauge-invariant site-wise norm 
$\rho^{(k)}(\vx,t)$ averaged over an ensemble of representative gauge 
configurations. The locality of the resulting field is seen as the field
takes large values close to the point where the source was anchored.

Smeared quark fields, $\chi$ are now given in terms of bare fields, $\psi$ by
\begin{align}
  \chi_\alpha^a(\vx,t) = \Box^{ab}(\vx,\vy;t) \psi_\alpha^b(\vy,t), 
\end{align}
where $\vx,\vy$ label spatial sites, $a,b$ color components and $\alpha$ the 
Dirac spinor index. Effective hadron creation operators are made from 
smeared quarks so for example an interpolating field for meson $M$ with 
momentum $\vp$ is defined by $\Gamma_M(\vp)$, a general gauge-covariant linear 
operator on time-slice $t$ which reads
\begin{align}
  \mathcal{O}_M^\dagger(t) = \bar\chi^a_\alpha(\vx,t) 
      \; \Gamma^{ab}_{M,\alpha\beta}(\vx,\vy;\vp, t) \;
                             \chi^b_\beta(\vy,t). 
\end{align}
Expanding this expression in terms of the unsmeared fields using the new 
localised distillation space basis vectors, $W(t)$ while  
suppressing spatial and color indices and keeping Dirac indices explicit 
gives 
\begin{align}
  \mathcal{O}^\dagger_M(t)
    &= \bar\psi_\alpha(t) \Box(t)
      \; \Gamma_{M,\alpha\beta}(\vp,t) \; \Box(t)
                             \psi_\beta(t) \nonumber\\
    &= \bar\psi_\alpha(t)\; W(t) W^\dagger(t) 
      \; \Gamma_{M,\alpha\beta}(\vp,t) \; W(t) W^\dagger(t)\;
                             \psi_\beta(t) \nonumber\\
    &= \bar\psi_\alpha(t) W(t) 
      \; \Phi_{M,\alpha\beta}(\vp,t) \; W^\dagger(t)
                             \psi_\beta(t),
\end{align}
where $\Phi_M(t)$ encodes the spatial, color and Dirac structure of operator
$\Gamma_M$. This can be decomposed efficiently into a sum of basis 
distillation-space matrices using
\begin{align}
  \Phi_{M,\alpha\beta}(\vp,t) &= \sum_k \phi^k(\vp,t) 
      S^k_{M,\alpha\beta} \nonumber\\
   \mbox{with } \phi^k(\vp,t) &= W^\dagger(t) \mathcal{D}^k (\vp,t) W(t), 
\end{align}
which factorises the operator into independent spinor matrices, $S^k_M$ 
generated using the relevant Clebsch-Gordan coefficients depending on the 
momentum and quantum numbers of meson $M$ and 
$\phi^k$ the distillation-space representations of 
$\mathcal{D}^k$, derivative-based operators with momentum $\vp$. 
These are called \textit{elementals}. 

\begin{figure*}[!thb]
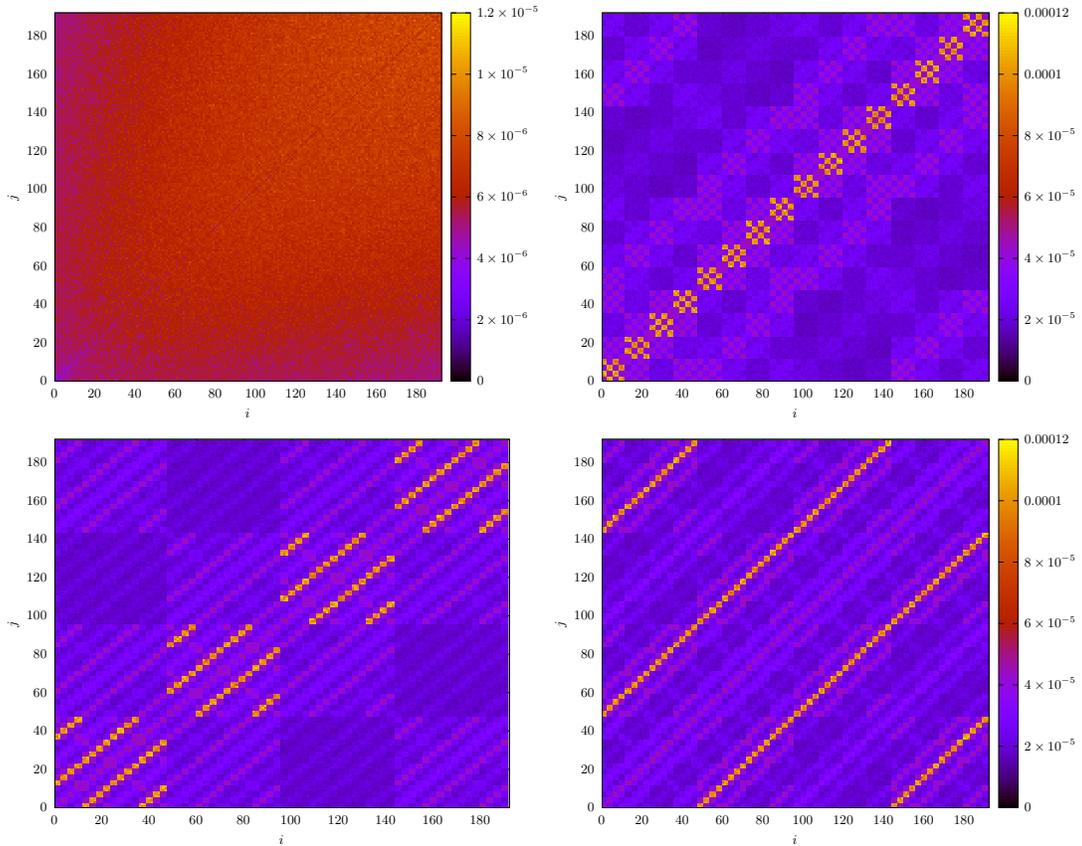

	\begin{subfigure}[l]{0.4\textwidth}
		\resizebox{\textwidth}{!}{
			\input{original_meson_abs_phases.d1.p000.tsv.matrix_pdf.tex}
		}
	\end{subfigure}
	\begin{subfigure}[l]{0.4\textwidth}
		\resizebox{\textwidth}{!}{
			\input{meson_abs_phases.d1.p000.tsv.matrix.tex}
		}
	\end{subfigure}
	\begin{subfigure}[l]{0.4\textwidth}
		\resizebox{\textwidth}{!}{
			\input{meson_abs_phases.d2.p000.tsv.matrix.tex}
		}
	\end{subfigure}
	\begin{subfigure}[l]{0.4\textwidth}
		\resizebox{\textwidth}{!}{
			\input{meson_abs_phases.d3.p000.tsv.matrix.tex}
		}
	\end{subfigure}
	\caption{Magnitudes of complex elements of meson elementals, $|\phi^M_{ij}(\vp=0,t)|$, constructed from non-rotated $v^{(i)}$ (top-left panel) and rotated $w^{(i)}$ (all other panels). Spatial derivatives in the x-direction (top left and right), y-direction (bottom-left) and z-direction (bottom-right) have been applied to the vectors as described in the text. The magnitudes of the matrix elements have been averaged over the ensemble of gauge configurations.}
	\label{fig:meson_elementals}
\end{figure*}

The magnitude of individual entries in these elementals depend strongly on the 
choice of basis and the aim of the new localised basis, $W$ is to introduce 
more structure to these matrices. 
As an example, for a gauge-invariant operator to create a meson with momentum 
$\vp$ 
made from smeared fields contracted without any derivatives applied, the 
elemental is evaluated using 
\begin{align}
  \phi^{(0)}_{ij}(\vp,t) = 
    \sum_{\vx,a} \;
     w^{(i)*}_a (\vx,t) \; e^{i\vp\cdot\vx} \; w^{(j)}_a(\vx,t). 
\end{align}
In this expression, $i$ and $j$ label the new local basis vectors for
distillation space. This corresponds to an inner product which has smaller magnitude for those
entries where the localisation anchor points are further apart and so $\phi$
has larger entries on its diagonal. Note if $\vp=0$, this matrix is the
identity (in any basis) due to the orthonormality of the spanning vectors.
Using Laplace eigenmodes as basis vectors however, the
matrix is dense for any non-zero momentum. Figure \ref{fig:meson_elementals}
illustrates the difference between these elementals in the two different bases
and the block-diagonal structure is seen clearly. 

\begin{figure*}[!thb]
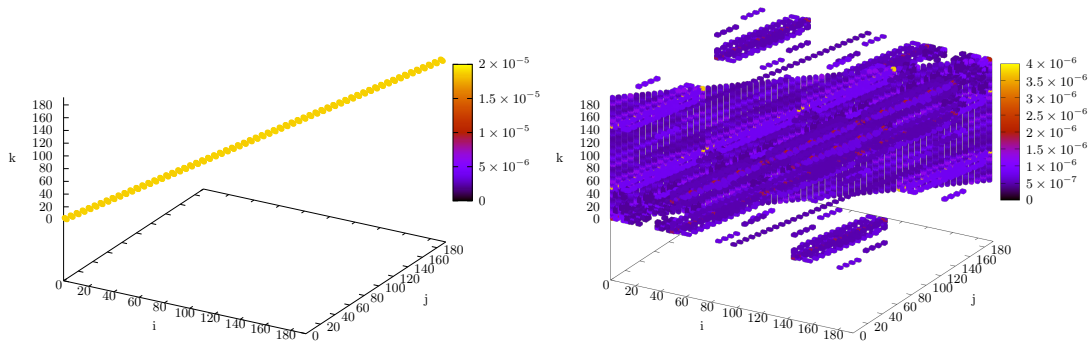

	\begin{subfigure}[l]{0.4\textwidth}
		\resizebox{\textwidth}{!}{
			\input{baryon_abs_phase.d...p000.tsv.scatter.tex}
		}
	\end{subfigure}
	\begin{subfigure}[l]{0.4\textwidth}
		\resizebox{\textwidth}{!}{
			\input{baryon_abs_phase.d..3.p000.tsv.scatter.tex}
		}
	\end{subfigure}
	\caption{Magnitudes of complex elements of baryon elementals, $|\phi^B_{ijk}(\vp=0,t)|$ constructed from rotated $w^{(i)}(t)$ with trivial spatial structure (left) and a spatial derivative in the z-direction applied to the last quark (right). Elements which are smaller than one-tenth of the maximum are suppressed in the plot for improved readability.}
	\label{fig:baryon_elementals}
\end{figure*}

For baryon creation operators, the benefits from localisation are more 
significant. Now the creation operator using distilled fields can be written
\begin{align}
\Phi^{B}_{\alpha \beta \gamma}(x,t) &= \sum_n \phi^{B,n}(x,t) S^
n_{\alpha \beta \gamma}\;, \nonumber \\
		\phi^{B,n}_{ijk}(x,t) &= \epsilon_{abc} \big(\mathcal{D}_n^1 
 {w}^{(i)}(t)\big)_x^a \big(\mathcal{D}_n^2 {w}^{(j)} (t)\big)_x^b 
  \big(\mathcal{D}_n^3 {w}^{(k)} (t)\big)_x^c.
		\label{eq:th:dist_baryon}
\end{align}
Even for the simplest baryon elemental containing no derivatives, all entries 
take non-zero 
values and in general they have similar magnitudes. If these entries are large 
only when the three anchor points are close then the number of large 
entries is reduced from ${\cal O}(n_D^3)$ to just ${\cal O}(n_D)$. This 
effect is seen in Figure~\ref{fig:baryon_elementals}.

Formally, the evaluation of the correlation function for these hadronic 
creation operators proceeds in exactly the same way as the original 
distillation algorithm. The correlation function for meson $M$ between a 
distilled source at time $t_0$ and sink at $t_1$ is written
\begin{align}
  C_M(t_0,t_1) &=  
\langle
 \mbox{Tr } \Phi(t_0) \tau(t_0,t_1) \Phi(t_1) \tau(t_1,t_0)
\rangle
\label{eq:meson_two_pt}
\end{align}
with $\tau(t_0,t_1)$ the perambulator, which describes quark 
propagation between the distillation spaces on time-slices $t_0$ and $t_1$, 
\begin{align}
  \tau_{\alpha\beta}(t_0,t_1) &=  
     W^\dagger(t_0) [M^{-1}]_{\alpha\beta}(t_0,t_1) W(t_1). 
\end{align}
In this expression, the Dirac indices $\alpha,\beta$ are explicitly given to 
emphasise this structure is inherited solely from the inverse of the lattice 
Dirac operator, $M$. In general, since the inverse Dirac operator appears in
the expression for the perambulator we will not assume using a localised basis 
for distillation space will lead to extra structure in the perambulator 
although this is worth further investigation. 
Following a similar procedure, two-point functions for the baryons can be 
expressed in terms of contractions of baryon elementals and perambulators. 

As a trace over products of matrices in distillation space, 
Eqn.~\ref{eq:meson_two_pt} expresses the correlation function as a large sum
over a number of distillation-space indices. In the following section, the
observation that localisation isolates large contributions in a small number 
of terms can be exploited to define stochastic representations with reduced 
variance compared to a similar scheme using the standard Laplace eigenvector 
basis for distillation space. 

\clearpage

\section{Stochastic computation of correlation functions \label{sec:sum}}
Consider $\sigma$, the sum 
of a set $\{a_1,a_2,\dots a_n\}$ of complex numbers. This sum can be estimated 
stochastically. The {\em Hansen-Hurwitz} estimator \cite{hansen1943} is a random number
defined by 
\begin{align}
  A = \left\{ 
    \begin{array}{ccc} 
      a_1/p_1 & \mbox{ with probability } & p_1,\\
      a_2/p_2 & \mbox{ with probability } & p_2,\\
         & \vdots & \\
      a_n/p_n & \mbox{ with probability } & p_n.
    \end{array}
      \right.
\end{align}
$\EX[A] = \sigma$ follows immediately from the definition for any set of 
normalised probabilities, $0<p_k\le 1$ with $\sum_k p_k=1$. 
The variance of the estimator is 
\begin{align}
	\VAR [A] = 
   \sum_{k=1}^{n} p_k \left|\frac{a_k}{p_k} - \sigma\right|^2 \;,
\end{align}
and the set of probabilities that minimises this variance is found to be 
\begin{align}
  \tilde{p}_k = \frac{|a_k|}{\sum_{j=1}^n |a_j|}. 
\end{align}
This choice implies an importance sampling where elements with large magnitude 
are more likely to be selected. With $p_k=\tilde{p}_k$ and writing 
$a_k=|a_k|e^{i\phi_k}$, the variance of the Hansen-Hurwitz estimator becomes
\begin{align}
	\VAR [A] = \sum_{k=1}^{n} \tilde{p}_k \left|  \sum_{j=1}^n |a_j| (e^{i \phi_k} - e^{i \phi_j})  \right|^2\;.
\end{align}
The variance is zero for the optimal sampling probabilities if
all entries in the sum have the same complex phase. The computation of 
hadronic two-point functions using distillation gives examples of large sums 
over complex variables and with the structure provided by the sparse basis 
introduced above, a simple importance sampling method might provide an 
efficient approach to the evaluation of these correlation functions.

\subsection{Estimating hadron correlation functions
       \label{subsec:sampled_correlators} } 
 Correlation functions between operators constructed in distillation 
spaces on two time-slices are a tensor contraction involving perambulators and 
source operators projected into the distillation space on a given time slice. 

To start with the simplest example, consider again a 
connected meson correlation function as seen in Eqn~\ref{eq:meson_two_pt}, and 
expand its evaluation on a single gauge-field background as
an explicit sum over sets of indices; 
\begin{align}
  c_{(AB)}(t_0,t_1) = 
     \sum_{abcd} \Phi^{(A)}_{ab}(t_0) \tau_{bc}(t_0,t_1) \Phi^{(B)}_{cd}(t_1) 
     \tau_{da}(t_1,t_0). 
     \label{eq:meson_contraction}
\end{align}
In this expression, each composite $\{a,b,c,d\}$ indicates a set of indices 
spanning a spin and distillation-space label to be summed over. As seen in 
Sec.~\ref{sec:sum}, the Hansen-Hurwitz estimator of this sum is defined by 
assigning a 
probability $p_{abcd}$ to every term and the importance sampling optimum is 
found by setting 
\begin{align}
  & \tilde p_{abcd}(t_0,t_1) =  \nonumber \\
      &\frac{|\Phi^{(A)}_{ab}(t_0) \tau_{bc}(t_0,t_1) 
             \Phi^{(B)}_{cd}(t_1) \tau_{da}(t_1,t_0) |}
 {\sum_{abcd} 
     |\Phi^{(A)}_{ab}(t_0) \tau_{bc}(t_0,t_1) 
             \Phi^{(B)}_{cd}(t_1) \tau_{da}(t_1,t_0) |}. 
\end{align}
Fully realising this optimum is impractical. To obtain these probabilities 
we would need to compute the full contraction, defeating the purpose of the 
sampling. Another difficulty is that the perambulator in the definition would 
couple the sampling of elements on distinct time-slices and prevent any data 
reuse. A simpler implementation is needed that captures the broad features 
of the importance sampling method, exploits the sparsity of the localised
basis and can be reused in a flexible way. 

One simplification is to define a sampling over the indices in the source 
and sink operators independently, which ignores structure in the perambulators 
and does not depend on the time index of the creation operators. By averaging over time slices and gauge configurations, we can significantly reduce the sensitivity of the sampling probabilities to fluctuations of the gauge field. 
Our final choice for the sampling probabilities reads
\begin{equation}
\begin{aligned} 
	\tilde{p}^{AB}_{abcd} &= 
\underbrace{\left\langle \frac{1}{N_t} \sum_{t=0}^{N_t} \frac{|\Phi^{(A)}_{ab}(t)|} {\sum_{ij} |\Phi^{(A)}_{ij}(t)|}
     \right\rangle}_{\tilde{p}^{(A)}_{ab} }\\ &\times
\underbrace{ \left\langle \frac{1}{N_t} \sum_{t=0}^{N_t} \frac{|\Phi^{(B)}_{cd}(t)|} {\sum_{kl} |\Phi^{(B)}_{kl}(t)|}
     \right\rangle}_{\tilde{p}^{(B)}_{cd} } \;,
\end{aligned} 
\end{equation}
where $\langle \ldots \rangle$ denotes the ensemble average.
The factorisation means the two pairs of indices $\{a,b\}$ and $\{c,d\}$ are 
drawn independently and the joint probability of all indices factorises so 
these samples can be reused in computations involving different time-slices 
for the source and sink. To be close to the optimum, the modulus of matrix 
entries in the perambulator must depend weakly on their location within the 
matrix and this dense structure is observed approximately for large 
time-separations. 
\begin{algorithm}[H]
\caption{Meson connected two-point correlation function 
   (Hansen-Hurwitz with replacement)} \label{alg1}
\begin{algorithmic}
\STATE Draw a random index pair $\{a,b\}$ with probability $\tilde p^{A}_{ab}$
\STATE Draw a random index pair $\{c,d\}$ with probability $\tilde p^{B}_{cd}$
\RETURN $\Phi^{(A)}_{ab}(t_0) \tau^{\vphantom (}_{bc}(t_0,t_1) 
    \Phi^{(B)}_{cd}(t_1) \tau^{\vphantom (}_{da}(t_1,t_0) / 
      (\tilde p^{(A)}_{ab} \tilde p^{(B)}_{cd})$
\end{algorithmic}
\end{algorithm}

The discussion to this point describes evaluation via sampling of a 
meson connected two-point correlation function. These ideas generalise to a 
broad class of correlation functions relevant for hadron physics. In the 
distillation framework, correlation functions are 
tensor contractions over sets of indices on the source and sink time-slices 
with the indices coupled via perambulators. Their evaluation reduces to a sum 
over the distillation-space and spin indices and so the Hansen-Hurwitz method,
combined with the sparse basis can be broadly applied. In particular, the 
advantage for contractions involving baryons, tetraquarks and beyond is 
potentially greater since these evaluations involve deeper nested sums. The
evaluation of correlation functions involving baryons is described in the next
section, where a hierarchy of efficient partial sums is needed. 

\subsection{Implementation of sparse tensor contractions
       \label{sec:tensor_contract} }

The term-by-term computation of the contraction performed in algorithm~\ref{alg1} is inefficient since the necessary complex-number multiplications scale like $\mathcal{O}(\nD^{2d})$ in the large-sample limit. However this can be reduced to $\mathcal{O}(\nD^{d+1})$ if the contractions are performed sequentially. In the case of the meson connected two-point correlation function we compute the \emph{temporary} tensors
\begin{equation}
	\begin{aligned}
		\Phi^{(A,1)}_{ac}(t_0, t_1) &= \sum_b \Phi^{(A)}_{ab}(t_0) \tau^{\vphantom (}_{bc}(t_0,t_1)\;,\\
		\Phi^{(A,2)}_{dc}(t_0, t_1) &= \sum_a \tau^{\vphantom (}_{da}(t_1,t_0) \Phi^{(A,1)}_{ac}
	\end{aligned}
	\label{eq:temporaries}
\end{equation}
and subsequently the full contraction
\begin{equation}
	c_{(AB)}(t_0,t_1) = \sum_{cd} \Phi^{(A,2)}_{dc}(t_0, t_1) \Phi^{(B)}_{cd}(t_1)\;.
\end{equation}
This computation can be combined with sampling in the following way.
First construct two vectors $\vs^{(A)}$ and $\vs^{(B)}$ of dimension $n_s$ with entries
\begin{equation}
	\begin{aligned}
		s^{(A)}_i &= (a,b)\; \text{with probability } \tilde{p}^A_{ab}\;,\\
		s^{(B)}_i &= (c,d)\; \text{with probability } \tilde{p}^B_{cd}\;,\\
	\end{aligned}
\end{equation}
called samples.
Now define the index-pair multiplicities in sample $\vs$ by
\begin{equation}
	m_{\vs}(a,b) = \sum_{i=1}^{n_s} \delta_{s_{i,1}a} \delta_{s_{i,2}b}
\end{equation}
where $s_{i,1}$ ($s_{i,2}$) references the first (second) component of the $i$th entry in sample $\vs$.
It will also be useful to define the indicator function
\begin{equation}
	I_{\vs}(a,b) = \Theta\left(m_{\vs}(a,b) - 1\right)
\end{equation}
and \emph{marginal} indicator function
\begin{equation}
	\label{eq:indi}
	I_{\vs}(\cdot,b) = \Theta\left(\sum_{a=1}^{\nD} m_{\vs}(a,b) - 1\right)
\end{equation}
(analogously $I_{\vs}( a, \cdot )$), where $\Theta(\ldots)$ is the Heaviside step function.
Now construct the weighted source and sink tensors 
\begin{equation}
	\begin{aligned}
		 \tilde{\Phi}^{(A)}_{ab} &= \frac{1}{ \tilde{p}^A_{ab}} m_{\vs^{(A)}}(a,b) \cdot \Phi^{(A)}_{ab}\; \text{and}\\
		 \tilde{\Phi}^{(B)}_{cd} &= \frac{1}{ \tilde{p}^B_{cd}} m_{\vs^{(B)}}(c,d) \cdot \Phi^{(B)}_{cd}\;.
	\end{aligned}
\end{equation}
To construct the first sparse temporary define the sets
\begin{equation}
	\begin{aligned}
		X^{(A)} &= \set{ (a,b) | a, b = 1, \ldots, \nD;  I_{\vs^{(A)}}(a,b) = 1 }\; \text{and} \\
		X^{(B,1)} &= \set{ c | c = 1, \ldots, \nD;  I_{\vs^{(B)}}(c,\cdot) = 1 }\;.
	\end{aligned}
\end{equation}
The following algorithm then computes the first sparse temporary from the samples.
%
%
\begin{algorithm}[H]
	\caption{Construction of first sparse temporary tensor from samples} \label{alg2}
	\begin{algorithmic}
		\STATE Initialize $\hat{\Phi}^{(A,1)}_{ab}(t_0, t_1) = 0$ for all $a$, $b \in 1, \ldots, \nD$
		\FOR{$(a,b) \in X^{(A)}$}
		\FOR{$c \in X^{(B,1)}$}
		\STATE $\hat{\Phi}^{(A,1)}_{ac}(t_0, t_1) = \tilde{\Phi}^{(A)}_{ab}(t_0) \tau^{\vphantom (}_{bc}(t_0,t_1)$
		\ENDFOR
		\ENDFOR
		\RETURN $\hat{\Phi}^{(A,1)}(t_0, t_1)$
	\end{algorithmic}
\end{algorithm}
This algorithm manifestly scales like $\mathcal{O}( \left| X^{(A)} \right| \cdot \left| X^{(B,1)} \right|)$.
Analogously, for the second sparse temporary we define the sets
\begin{equation}
	\begin{aligned}
		X^{(B)} &= \set{ (c,d) | c, d = 1, \ldots, \nD;  I_{\vs^{(B)}}(c,d) = 1 }\; \text{and} \\
		X^{(A,1)} &= \set{ a | a = 1, \ldots, \nD;  I_{\vs^{(A)}}(\cdot, b) = 1 }\;,
	\end{aligned}
\end{equation}
and execute the sparse contraction as per the following algorithm.
\begin{algorithm}[H]
	\caption{Construction of second sparse temporary tensor from samples} \label{alg3}
	\begin{algorithmic}
		\STATE Initialize $\hat{\Phi}^{(A,2)}_{dc}(t_0, t_1) = 0$ for all $c$, $d \in 1, \ldots, \nD$
		\FOR{$(c,d) \in X^{(B)}$}
		\FOR{$a \in X^{(A,1)}$}
		\STATE $\hat{\Phi}^{(A,2)}_{dc}(t_0, t_1) =  \tau^{\vphantom (}_{da}(t_1,t_0) \tilde{\Phi}^{(A,1)}_{ac}(t_0, t_1)$
		\ENDFOR
		\ENDFOR
		\RETURN $\hat{\Phi}^{(A,2)}(t_0, t_1)$
	\end{algorithmic}
\end{algorithm}
Lastly the correlator can be computed from the second sparse temporary and the weighted sink.
\begin{algorithm}[H]
	\caption{Computation of meson connected two-point correlation function from sparse temporaries} \label{alg4}
	\begin{algorithmic}
		\STATE Initialize $\hat{c}_{(AB)}(t_0,t_1) = 0$
		\FOR{$(c,d) \in X^{(B)}$}
		\STATE $\hat{c}_{(AB)}(t_0,t_1) \pluseq \hat{\Phi}^{(A,2)}_{dc}(t_0, t_1) \tilde{\Phi}^{(B)}_{cd}(t_1)$
		\ENDFOR
		\RETURN $\hat{c}_{(AB)}(t_0,t_1)/n_s^2$
	\end{algorithmic}
\end{algorithm}

In the case of baryon operators such as the ones we will discuss in the next section, the samples are vectors of index triplets. 
For example, we may compute the contraction
\begin{align*}
	&c_{(AB)}(t_0,t_1) =\\ &\sum_{\substack{abc\\def}} \Phi^{(A)}_{abc}(t_0) \tau^{(1)}_{ad}(t_0,t_1)  \tau^{(2)}_{be}(t_0,t_1) \tau^{(3)(}_{cf}(t_0,t_1)
	\Phi^{(B)}_{def}(t_1) 
\end{align*}
where the perambulator superscripts indicate that these may be of different quark flavours.
Given weighted source and sink tensors $\tilde{\Phi}^{(A)}_{abc}$ and $\tilde{\Phi}^{(B)}_{def}$, to construct the first temporary we define the sets
\begin{equation}
	\begin{aligned}
		X^{(A)} &= \set{ (a,b,c) | a, b, c = 1, \ldots, \nD;  I_{\vs^{(A)}}(a,b,c) = 1 }\; \\
		&\text{and }X^{(B,2)} = \set{ d | d = 1, \ldots, \nD;  I_{\vs^{(B)}}(d,\cdot, \cdot) = 1 }\;.
	\end{aligned}
\end{equation}
where $I_{\vs}(a,b,c)$ is the indicator function for the triplet $(a,b,c)$ in the sample $\vs$ and
\begin{equation}
	I_{\vs}(d,\cdot,\cdot) = \Theta \left(  \sum_{e,f=1}^{\nD} I_{\vs}(d,e,f) - 1 \right)
	\label{eq:indi_baryon}
\end{equation}
is the marginal indicator function for the first index. Then algorithm~\ref{alg5} computes the first sparse temporary.
\begin{algorithm}[H]
	\caption{Construction of first sparse temporary tensor for a baryon contraction}
	\label{alg5}
	\begin{algorithmic}
		\STATE Initialize $\hat{\Phi}^{(A,1)}_{abc}(t_0, t_1) = 0$ for all $a$, $b$, $c \in 1, \ldots, \nD$
		\FOR{$(a,b,c) \in X^{(A)}$}
		\FOR{$d \in X^{(B,2)}$}
		\STATE $\hat{\Phi}^{(A,1)}_{dbc}(t_0, t_1) = \tilde{\Phi}^{(A)}_{abc}(t_0) \tau^{(1)}_{ad}(t_0,t_1)$
		\ENDFOR
		\ENDFOR
		\RETURN $\hat{\Phi}^{(A,1)}(t_0, t_1)$
	\end{algorithmic}
\end{algorithm}
The other temporaries and full contraction follow the same pattern.
This procedure generalises to arbitrary tensor contractions.
For connected two-point correlation functions the number of temporaries is equal to the number of open distillation-space indices on the source and sink tensors. The sets $X^{(A,r-1)}$ or $X^{(B,d-r)}$, which are inputs for the algorithm that builds the $r$th temporary, are constructed analogously using generalisations of the indicator functions defined in equation~\ref{eq:indi_baryon}. The most general form must allow for permutations $P$ of the quark lines and can be written
\begin{equation}
	\begin{aligned}
	I^{k,P}_{\vs}&(i_{P(k+1)}, \ldots, i_{P(d)}) =\\ &\Theta \left(  \sum_{i_{P(1)},\ldots, i_{P(k)}=1}^{\nD} I_{\vs}(i_1, \ldots, i_d) - 1 \right)
	\end{aligned}
\end{equation}
where $d$ is the dimension of the source tensor and $k < d$.

\section{
   Results from first tests.  \label{sec:application} } 

As a test case for this new contraction method we compute the single-baryon $c_{(N \rightarrow N)}(t) = \braket{0|\mathcal{O}_N(t)  \mathcal{O}^\dagger_N(0)|0}$ correlator in the $I(J)^P = 1/2(1/2)^+$ channel. We also compute the correlators $c_{(\Delta \rightarrow \Delta)}(t) = \braket{0|\mathcal{O}_\Delta(t)  \mathcal{O}^\dagger_\Delta(0)|0}$ and $c_{(N\pi \rightarrow \Delta)}(t) = \braket{0|\mathcal{O}_\Delta(t)  \mathcal{O}^\dagger_{N\pi} (0)|0}$, where the operators interpolate at-rest single-hadron and two-hadron $I(J)^P = 3/2(1/2)^-$ states.

\begin{figure*}[!thb]
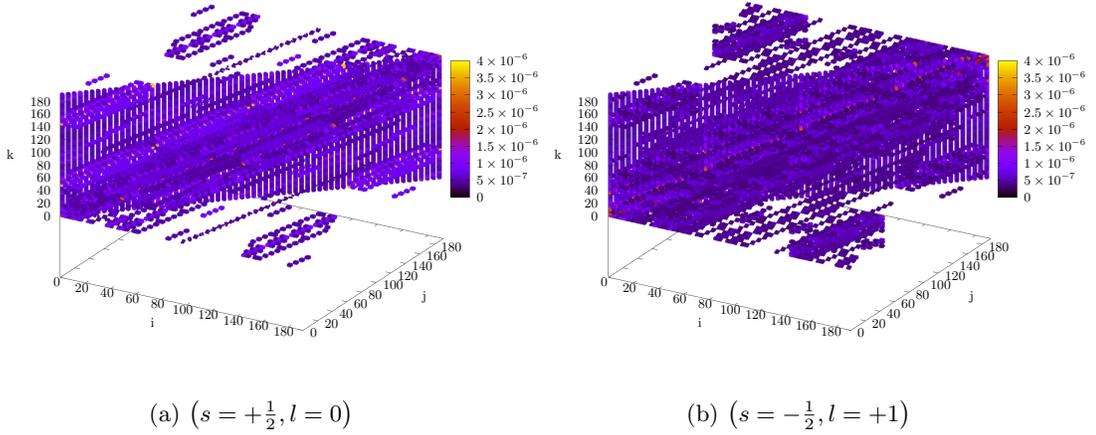

	\begin{subfigure}[l]{0.4\textwidth}
		\resizebox{\textwidth}{!}{
			\input{delta_s001_pdf.dat1000a.scatter.tex}
		}
		\subcaption{$\left(s=+\tfrac{1}{2},l=0\right)$}
	\end{subfigure}
	\begin{subfigure}[l]{0.4\textwidth}
		\resizebox{\textwidth}{!}{
			\input{delta_s110_pdf.dat1000a.scatter.tex}
		}
		\subcaption{$\left(s=-\tfrac{1}{2}, l=+1\right)$}
	\end{subfigure}
	\vspace{0.5cm}
	\caption{Sampling probabilities of the two spin components of $\mathcal{O}^{J^P=\tfrac{1}{2}^-}_\Delta$ that are distinct in terms of their weight densities.
          Panel (a) shows the weights with a $D_z$ angular momentum structure ($l=0$) and
          panel (b) shows weights with the angular momentum structure $D_x + i D_y$ ($l=+1$).
          The other four weight tensors can be obtained by index permutations. Only tensor entries larger than one tenth of the maximum are shown.}
	\label{fig:delta_weights}
\end{figure*}

In the construction of the baryon operators we follow the procedure described in detail in Ref.~\cite{Edwards:2011jj}. As explained there, in any baryon operator the contraction of the colour indices with the totally-antisymmetric tensor requires the remaining quantum numbers to be symmetric under the interchange of two quarks for the operator to obey the Pauli principle. For simplicity and since all our calculations will be at zero overall momentum, we use two-component Pauli spinors in our constructions. Following the notation in Ref.~\cite{Edwards:2011jj}, our nucleon operator is denoted as
\begin{equation}
	\mathcal{O}^{J^P=\tfrac{1}{2}^+}_N = \left(N_M \otimes (\tfrac{1}{2}^+)_M \otimes \mathbf{1}_{L=0,S} \right)^{J^P=\tfrac{1}{2}^+}
\label{eq:nucop}
\end{equation}
meaning that this operator is mixed-symmetric in spin and flavour and
symmetric (and trivial) in space, giving an overall symmetric
combination of quantum numbers.
Since the simplest nucleon operator has no derivatives, its distillation space projection is sparse. Only tensor components corresponding to the same coarse grid site have large magnitudes. These are confined to three-by-three-by-three cubes on the body diagonal of the tensor in index space (see left-hand panel of Fig.~\ref{fig:baryon_elementals}). 
 Our $\Delta$ operator has the construction
\begin{equation}
	\mathcal{O}^{J^P=\tfrac{1}{2}^-}_\Delta = \left(\Delta_S \otimes (\tfrac{1}{2}^+)_M \otimes D^{[1]}_{L=1,M} \right)^{J^P=\tfrac{1}{2}^-}
\label{eq:deltaop}
\end{equation}
meaning it is flavour-symmetric under quark exchange and
mixed-symmetric in spin and derivatives, again giving an
overall-symmetric combination. The angular momenta of the spinors
and derivatives are combined using the standard Clebsch-Gordan formula of
$SU(2)$ in order to obtain operators with good $J$ in the
continuum,
\begin{equation}
  \big|\, [J, M] \big\rangle  = \sum_{s,l}  \big|\, [S,s] \big\rangle \otimes \big|\, [L, l] \big\rangle
\big\langle S s; L l \big| J M \big\rangle.
\end{equation}
For the $\Delta$ with total spin $J=M=\tfrac{1}{2}$, spin $S=\tfrac{1}{2}$ and angular momenta of the derivatives $L=1$, there are two terms contributing in the sum with the pairs $\left(s=+\tfrac{1}{2},l=0\right)$ and $\left(s=-\tfrac{1}{2}, l=+1\right)$. Thus, for the baryon elemental construction from Eqn.~\ref{eq:th:dist_baryon}, there are two spatial structures, shown in Fig.~\ref{fig:delta_weights}. The $l=0$ term involves a single gauge-covariant derivative $D_z$ in the spatial $z$ direction, while the $l=+1$ term involves the circular basis construction of $D_x + i D_y$, thus derivatives in the spatial $x$ and $y$ directions. This latter term is therefore denser in distillation space. Finally, both the nucleon and $\Delta$ operator subduce trivially into the $G_1$ lattice irreducible representation (irrep), and there are no additional reweightings of baryon elemental structures.

For the $I=1$ pion operator we use the spatially trivial construction
\begin{equation}
	\mathcal{O}^{J^P=0^-}_{\pi}(\vp = 0, t) = \sum_{\vx \in \Lambda_3} \bar{u}(\vx, t) \gamma_5 d(\vx, t)
\end{equation}
which leads to
\begin{equation}
	\begin{aligned}
		\Phi^{\pi}_{\alpha \beta}(\vp = 0, t) &= W^\dagger(t) W(t) \cdot (\gamma_5)_{\alpha \beta}\\
		&= \mathbf{1}  \cdot (\gamma_5)_{\alpha \beta} \;.
	\end{aligned}
\end{equation}
The two-hadron $N\pi$ operator is built from the operator product of $O^{J^P=\tfrac{1}{2}^+}_N$ and $O^{J^P=0^-}_\pi$, each projected onto zero spatial momentum, and projected to definite spin and isospin. Since we only compute the rows of maximum $J_3$ and $I_3$ we have
\begin{equation}
	\braket{\tfrac{1}{2}, \tfrac{1}{2}; 0,0 | \tfrac{1}{2}, \tfrac{1}{2}}_J = 1
\end{equation}
and
\begin{equation}
	\braket{\tfrac{1}{2}, \tfrac{1}{2}; 1,1 | \tfrac{3}{2}, \tfrac{3}{2}}_I = 1\;,
\end{equation}
which simplifies the construction.
The resulting operator $O^{J^P=\tfrac{1}{2}^-}_{N\pi}$ subduces one-to-one in the $G_1$ lattice irrep.

\begin{figure*}[!thb]
	\begin{subfigure}[l]{0.4\textwidth}
		\includegraphics[width=\textwidth]{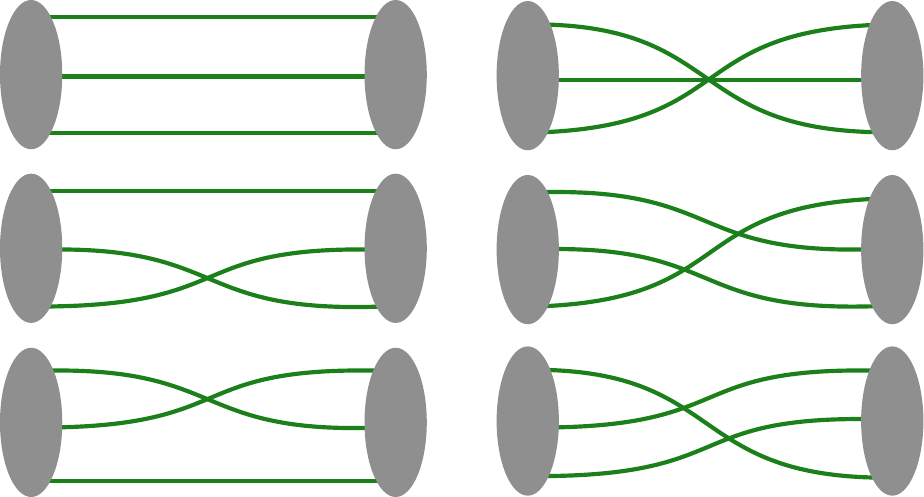}
		\subcaption{$c_{(\Delta \rightarrow \Delta)}$}
		\label{fig:wick-topologies:delta}
	\end{subfigure}
	\hspace{2cm}
	\begin{subfigure}[l]{0.4\textwidth}
		\includegraphics[width=\textwidth]{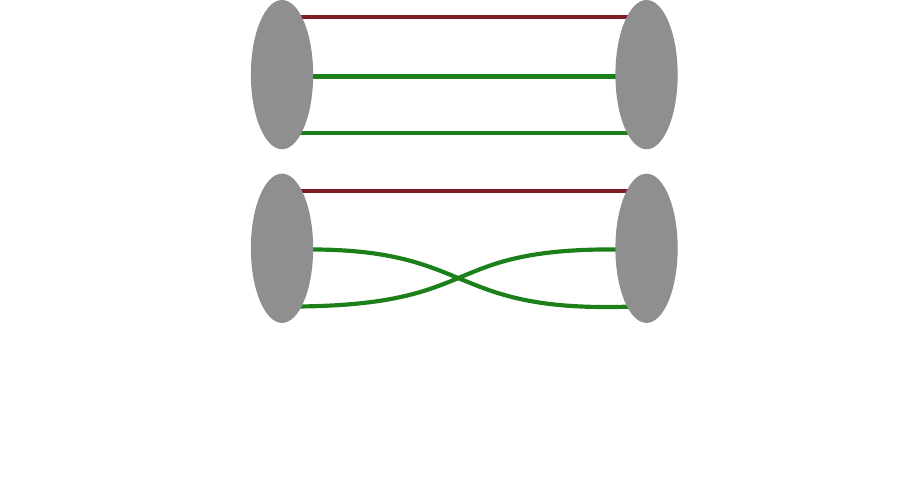}
		\subcaption{$c_{(N \rightarrow N)}$}
		\label{fig:wick-topologies:N}
	\end{subfigure}
	\begin{subfigure}[l]{0.4\textwidth}
		\vspace{1cm}
		\includegraphics[width=\textwidth]{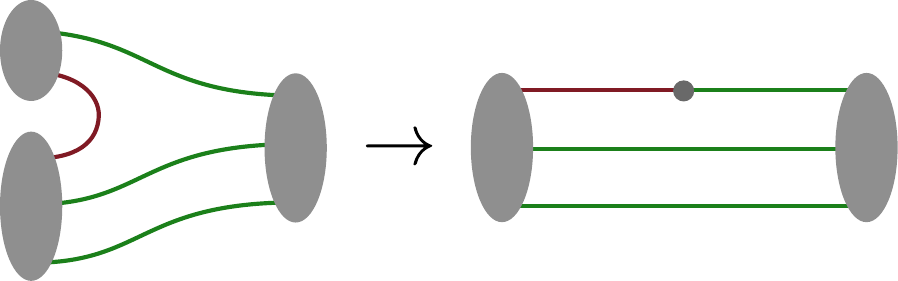}
		\subcaption{$c_{(N \pi \rightarrow \Delta)}$ (+ permutations as in panel~\ref{fig:wick-topologies:delta})}
		\label{fig:wick-topologies:Npi}
	\end{subfigure}
	\vspace{1cm}
	\caption{Topologies of diagrams contributing to the correlators discussed in section \ref{sec:application}. Green and red lines correspond to two distinct light flavours of quark. The line with a dark circle on the bottom panel represents a pion contracted with quark propagators on both sides.}
	\label{fig:wick-topologies}
\end{figure*}

The topologies of the possible diagrams appearing in $c_{(N \rightarrow N)}$, $c_{(\Delta \rightarrow \Delta)}$ and $c_{(N\pi \rightarrow \Delta)}$ are shown in Fig.~\ref{fig:wick-topologies}. Note that $c_{(N\pi \rightarrow \Delta)}$ has a quark-annihilation line at the source. Since the pion at zero momentum is trivial in distillation space, we can simplify the calculation by defining a pion perambulator, given by
\begin{equation}
	\begin{aligned}
		\tilde{\tau}_{\alpha \delta}(t,0) &= \tau_{\alpha \beta}(t,0) \Phi^{\pi}_{\beta \gamma}(\vp = 0, 0) \tau_{\gamma \delta}(0,0)\\
		&=  \tau_{\alpha \beta}(t,0) (\gamma_5)_{\beta \gamma} \tau_{\gamma \delta}(0,0)\;.
	\end{aligned}
\end{equation}
With this definition the topologies of the diagrams appearing in $c_{(N\pi \rightarrow \Delta)}$ are identical to those in $c_{(\Delta \rightarrow \Delta)}$, but in this case source and sink operators differ. This is illustrated for the first of the six diagrams in Fig.~\ref{fig:wick-topologies:Npi}. The pion perambulator is represented by a line adorned with a filled circle.

For each diagram topology all quark-spin combinations corresponding to $S=\frac{1}{2}$ need to be computed. This amounts to 36 individual spin contributions to each of the six diagram topologies of $c_{(\Delta \rightarrow \Delta)}$. However, after performing the spin sum, the diagrams in Fig.~\ref{fig:wick-topologies:delta} are all identical such that we have to compute 36 distillation-space sums overall for the $\Delta$-to-$\Delta$ correlator. 
In the case of the nucleon, due to the flavour structure the operator can be constructed in a way that only two quark-spin configurations are needed on each end of the correlator.
Together with the two distinct diagram topologies shown in Fig.~\ref{fig:wick-topologies:N}, this amounts to eight full contractions in distillation-space. 
For $c_{(N \pi \rightarrow \Delta)}$, the six diagram topologies are again identical after summing over spins, such that a total of twelve distillation-space contractions are required considering the two spin-configurations of the $N\pi$ operator and six of the $\Delta$ operator. Note that there are internal spin-sums appearing in the contractions of the operators with the perambulators that we need to keep track of.

\begin{figure*}[htb!]
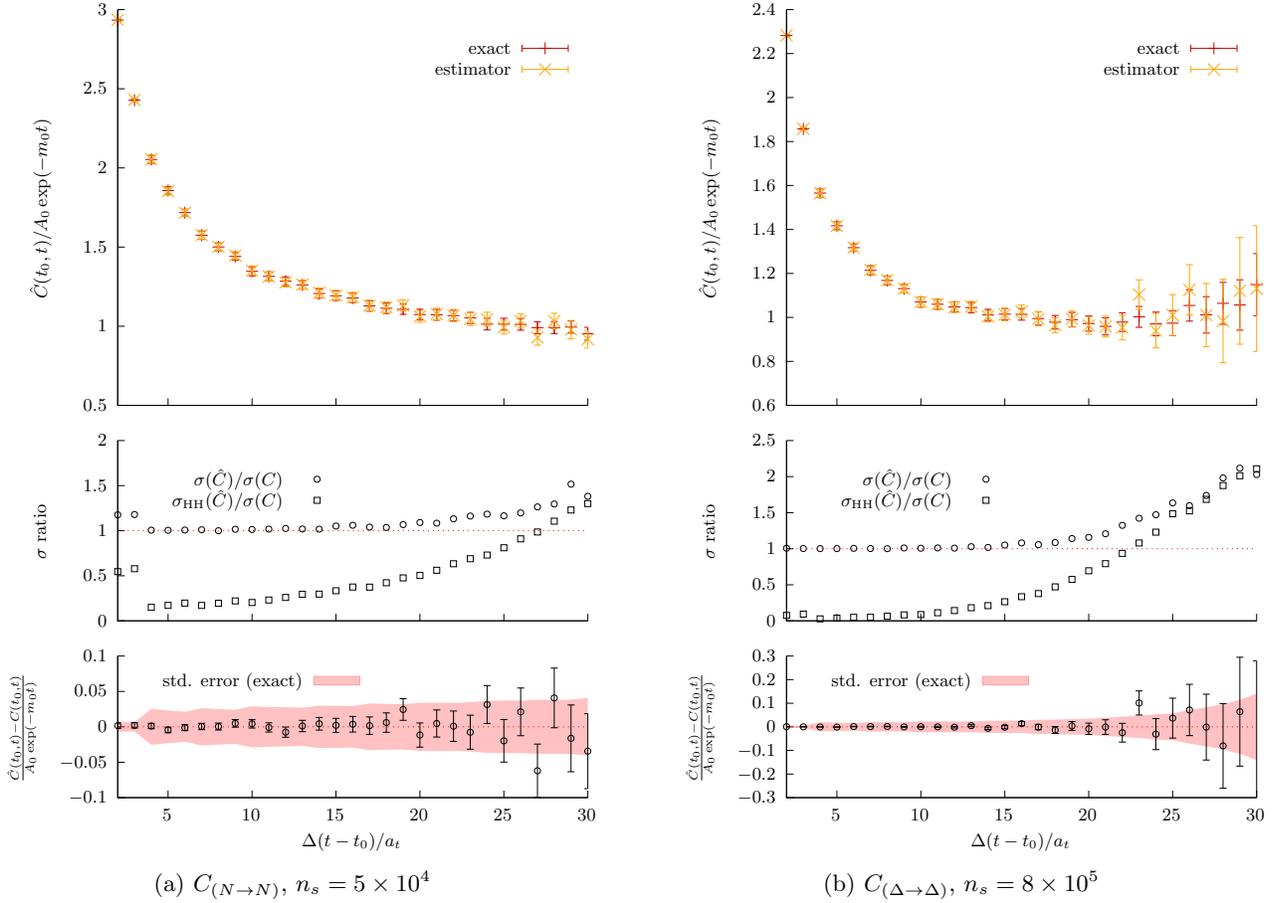

	\begin{subfigure}[l]{0.49\textwidth}
		\resizebox{\textwidth}{!}{
			\input{nn_corr_md_wr_over_exp_latex2.tex}
		}
		\subcaption{$C_{(N \rightarrow N)}$, $n_s = 5 \times 10^4$}
		\label{fig:corrs:nucl}
	\end{subfigure}
	\begin{subfigure}[l]{0.49\textwidth}
		\resizebox{\textwidth}{!}{
			\input{delta_delta_corr_wr_over_exp_latex3.tex}
		}
		\subcaption{$C_{(\Delta \rightarrow \Delta)}$, $n_s = 8 \times 10^5$}
		\label{fig:corrs:delta}
	\end{subfigure}
	\vspace{0.25cm}
\caption{$C_{(N \rightarrow N)}$ and $C_{(\Delta \rightarrow \Delta)}$ computed with exact and sampled distillation using sampling with replacement and the Hansen-Hurwitz estimator. The correlators are rescaled by the leading exponential obtained from a fit to the exact data. In each plot the middle panel shows the ratio of the standard error between the exact and the sampled correlator as well as an estimate of the contribution to the error that is solely due to the sampling procedure. The bottom panel shows the absolute difference between correlator values computed by the two methods divided by the leading exponential. The error bars in the bottom panels represent the sampling error only.}
\label{fig:corrs}
\end{figure*}

The computation of the correlator is carried out on an ensemble of 482 gauge configurations over 32 time-slices using a single time source.
The ensemble has been generated using a tree-level Symanzik-improved anisotropic action for the gauge fields and a tree-level, tadpole-improved Sheikholeslami-Wohlert action with $N_f = 2+1$ dynamical quark flavours for the fermion sector, where stout-smearing has been applied to the spatial gauge links~\cite{Morningstar_2004}.
On each gauge configuration and time-slice, and for each configuration of quark spins, we calculate the Hansen-Hurwitz estimate of the correlator, given by
\begin{equation}
	\hat{c}_{(AB)}(t_0, t_1) = \frac{1}{n_s^2} \sum_{\substack{(abc) \in s^{(A)} \\ (def) \in s^{(B)}}}  \frac{\left[c_{(AB)}(t_0, t_1)\right]_{abcdef}}{\tilde{p}^{(A)}_{abc} \times \tilde{p}^{(B)}_{def}} \;,
\end{equation}
from two samples $s^{(A)}$ and $s^{(B)}$ of lengths $n_s$ drawn from the distributions $\tilde{p}^{(A)}$ and $\tilde{p}^{(B)}$ respectively, where $A$ and $B$ are placeholders for the aforementioned operators.
For example, for the first diagram in Fig.~\ref{fig:wick-topologies:N} we compute
\begin{equation}
	\begin{aligned}
		&\left[c_{(N \rightarrow N)} (t_0, t_1)\right]_{abcdef} = \\
		&\frac{(\Phi^{(N)})^\dagger_{abc}(t_0) \tau^{\vphantom (}_{ad}(t_0,t_1) \tau^{\vphantom (}_{be}(t_0,t_1) \tau^{\vphantom (}_{cf}(t_0,t_1) 
			\Phi^{(N)}_{def}(t_1)}{\tilde{p}^{(N)}_{abc} \times \tilde{p}^{(N)}_{def}}
	\end{aligned}
\end{equation}
for all index tuples $(abc)$ and $(def)$ contained in two distinct samples drawn from $\tilde{p}^{(N)}$, where $\Phi^{(N)}(t)$ is the distillation-space projection of $\mathcal{O}^{J^P=\tfrac{1}{2}^+}_N$ on time-slice $t$ for the given set of quark spins.
An unbiased estimator for the variance of $\hat{c}_{(AB)}(t_0, t_1)$ is given by
\begin{equation}
	\begin{aligned}
		\hat{\sigma}_{\text{HH}}^2 &\left[ \hat{c}_{(AB)}(t_0, t_1) \right] = \frac{1}{n_s^2 (n_s^2 -1)} \sum_{\substack{(abc) \in s^{(A)} \\ (def) \in s^{(B)}}} \Biggl\{\\
		 &  \frac{\left|\left[c_{(AB)}(t_0, t_1)\right]_{abcdef}\right|^2}{\left(\tilde{p}^{(A)}_{abc} \times \tilde{p}^{(B)}_{def}\right)^2} - \left|\hat{c}_{(AB)}(t_0, t_1)\right|^2 \Biggr\}  \;
	\end{aligned}
\end{equation}
and computed in the same way as the correlator using the method presented in section~\ref{sec:tensor_contract}.
Since the nucleon is less dense in distillation space than the $\Delta$, $\hat{c}_{(N \rightarrow N)}$ can be computed with a smaller sample size than the correlators involving the $\Delta$ operator.
We choose a sample size $n_s = 5 \times 10^4$ for the former whereas $\hat{c}_{(\Delta \rightarrow \Delta)}$ and $\hat{c}_{(N\pi \rightarrow \Delta)}$ are sampled with $n_s = 8 \times 10^5$. In all cases we choose symmetric sample sizes for the two samples corresponding to the source and sink operators. Furthermore sampling is only applied in distillation space such that sums over Dirac-space are always computed exactly. Distillation-space samples are drawn individually for each set of Dirac components.
We also compute the exact contractions using full distillation in order to have a reference value. The results of these computations are shown in Fig.~\ref{fig:corrs}.
We denote the ensemble averages with capital letters, i.e. $C_{(AB)}(t_0, t) = \langle c_{(AB)}(t_0, t) \rangle$, and capital letters with hats denote ensemble averages of estimators. $\sigma_{\text{HH}}(\hat{C})$ denotes the ensemble average of the standard deviation of the Hansen-Hurwitz estimator whereas $\sigma(\hat{C})$ denotes the full standard deviation of the correlator including sampling and gauge noise. Ensemble quantities are estimated using jack-knife resampling.

For all three correlators we are able to obtain an unbiased estimate from the sampling procedure. We find that $\hat{C}_{(N \rightarrow N)}$ estimates the exact result with good precision across all time-slices. The additional noise introduced by the Hansen-Hurwitz sampling is sub-leading with respect to the gauge noise (see middle panel of Fig.~\ref{fig:corrs:nucl}).  $\hat{C}_{(\Delta \rightarrow \Delta)}$ has good precision up to $(t - t_0)/a_t \approx 22$, at which point a plateau has formed. $\hat{C}_{(N\pi \rightarrow \Delta)}$ is substantially more noisy and there is a significant contribution to the noise due to the Hansen-Hurwitz sampling even at early time-slices. We suspect that cancellations due to complex phases make the sampling less efficient and that this effect may be enhanced in the case of unequal source and sink operators.

\begin{figure*}[htb!]
	\begin{subfigure}[l]{0.5\textwidth}
		\resizebox{\textwidth}{!}{
			\input{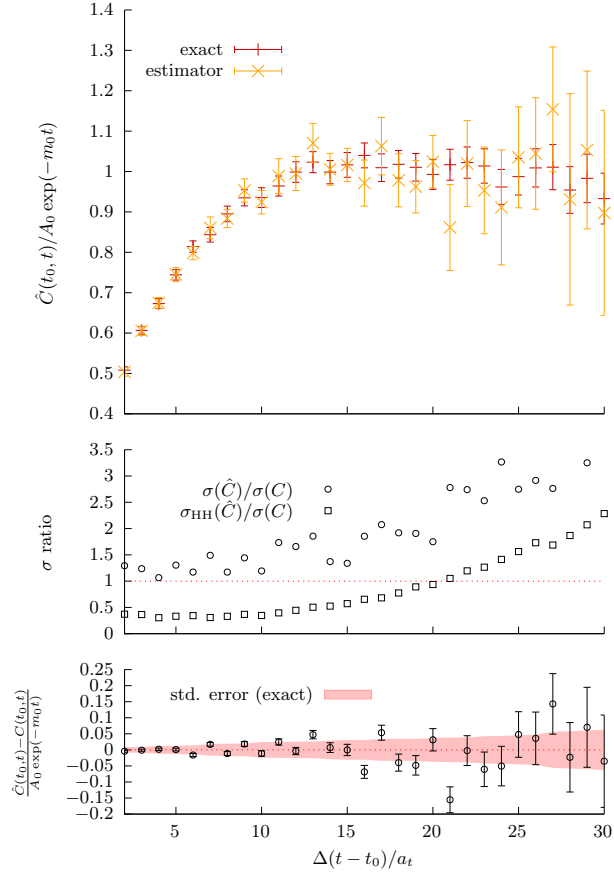}
		}
		\label{fig:corrs:npidelta}
	\end{subfigure}
	\vspace{0.25cm}
	\caption{Like figure \ref{fig:corrs} but showing $C_{(N\pi \rightarrow \Delta)}$ with  $n_s = 8 \times 10^5$.}
	\label{fig:corrs2}
\end{figure*}

\section{Discussion}
	\label{sec:discussion}
Having demonstrated the viability of the localized distillation method, we dedicate this section to some general observations regarding the efficiency and scaling of the algorithm. We do not attempt to make precise statements about the actual computer time but rather identify the conditions under which localized distillation can have substantial advantages over standard distillation.

\subsection{Efficiency of the method for the evaluation of baryon correlation 
  functions}
\label{subsec:baryon-efficiency}

For baryons, the most expensive part of the sampling method is the 
construction of the second temporary. To see this, recall the algorithm to 
construct the 
$r$th temporary scales like $\mathcal{O}( \left| X^{(A, r-1)} \right| \cdot \left| X^{(B,d-r)} \right|)$, where $X^{(A, k)}$ ($X^{(B, k)}$) is the set of distinct index 
tuples in the sample of $\mathcal{O}^A$ ($\mathcal{O}^B$) marginalised over the $k$ outermost (innermost) index-dimensions.
As a simple example, consider $\nD=2$, $d=3$. Assume for simplicity that the same sample of indices $\vs = \left[ (1,1,1), (1,2,1),  (2,2,1), (1,1,1), (1,2,2) \right]$ has been drawn for operators $\mathcal{O}^A$ and $\mathcal{O}^B$.
Since we can factor multiply-occurring indices into the weights, we
construct the set of unique indices $X^B = X^A \equiv X^{A,0} = \set{(1,1,1),
(1,2,1), (1,2,2), (2,2,1)}$. The first marginal set for operator $\mathcal{O}^A$ is then obtained by 
marginalising over the outermost index and given by $X^{(A,1)} = 
  \set{(1,1), (1,2), (2,2)}$. The second marginal set for $A$ is 
$X^{(A,2)} = \set{(1), (2)}$. For $\mathcal{O}^B$ one has to marginalise over the innermost indices, so $X^{(B,1)} = \set{(1,1), (2,1), (2,2)}$ and $X^{(B,2)} = \set{(1), (2)}$. Note that the maximum size of the sets
$X^{(A, k)}$ and $X^{(B, k)}$  is $\nD^{d-k}$. We can then define the \emph{sampling occupancy} of each 
set as $\rho^A_k = \left| X^{(A, k)} \right| / \nD^{d-k}$ and analogously $\rho^B_k$. In this example we 
would obtain $\rho^A_0 = \rho^B_0 = \frac{1}{2}$, $\rho^A_1 = \rho^B_1 = \frac{3}{4}$ and 
$\rho^A_2 = \rho^B_2 = 1$. Defining the cost coefficient $\zeta_r = \rho^A_{r-1} \cdot \rho^B_{d-r}$, the cost of the computation of the $r$th temporary is given by $\mathcal{O}(\zeta_r \nD^{d+1})$. In our example we would get $\zeta_1 =  \frac{1}{2}$, $\zeta_2 = \frac{9}{16}$ and $\zeta_3 = \frac{1}{2}$.
This illustrates that the cost coefficient pertaining to the second temporary is largest.
The fact that we assumed identical samples on both sides does not affect this point.
For a given sample size $n_s$ with replacement, the occupancies can be 
obtained directly from the weights. 
The expected value for the size of the set $X^A$ is 
\begin{equation}
	\EX\left[\left| X^{(A)} \right|\right] = \nD^d - \sum_{i_1, \ldots, i_d = 1}^{\nD} (1-\tilde{p}^{(A)}_{i_1, \ldots, i_d})^{n_s} \;.
\end{equation}
For the marginal sets we define the marginal probabilities
\begin{equation}
	\begin{aligned}
		\tilde{p}^{(A,k)}_{i_1, \ldots, i_{d-k}} = \sum_{i_{d-k+1}, \ldots, i_d = 1}^{\nD}  \tilde{p}^{(A)}_{i_1, \ldots, i_d}\;,
	\end{aligned}
\end{equation}
such that
\begin{equation}
	\EX\left[\left| X^{(A,k)} \right|\right] = \nD^{d-k} - \sum_{i_1, \ldots, i_{d-k} = 1}^{\nD} (1-\tilde{p}^{(A,k)}_{i_1, \ldots, i_{d-k}})^{n_s} \;.
\end{equation}
The marginal probabilities for operator $\mathcal{O}^B$ are constructed 
analogously (summing over the innermost indices). If $\mathcal{O}^A = \mathcal{O}^B$ and the operator has no derivatives (and is therefore anti-symmetric under index-permutations) they are equal.

\begin{figure*}[!thb]
	\begin{subfigure}[l]{0.45\textwidth}
		\resizebox{\textwidth}{!}{
			\input{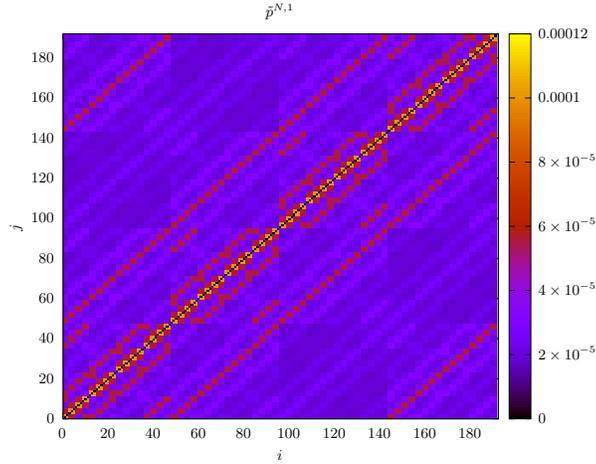}
		}
	\end{subfigure}
	\caption{The distribution of probability weights marginalised 
   over one index for the evaluation of the nucleon correlation function.}
	\label{fig:reduced_nucleon_weights}
\end{figure*}

\begin{figure*}[!thb]
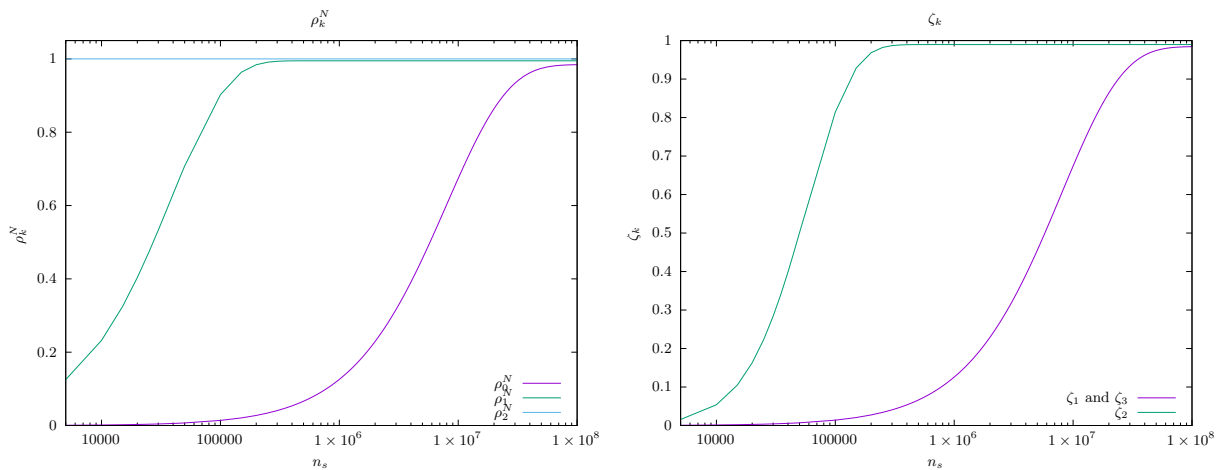

	\begin{subfigure}[l]{0.45\textwidth}
		\resizebox{\textwidth}{!}{
			\input{rhoNk.tex}
		}
	\end{subfigure}
	\begin{subfigure}[l]{0.45\textwidth}
		\resizebox{\textwidth}{!}{
			\input{rhoK.tex}
		}
	\end{subfigure}
	\caption{Expected index-set occupancies (left) and the cost factors of the three temporaries (right) of the nucleon operator}
	\label{fig:exp_N}
\end{figure*}

Fig.~\ref{fig:reduced_nucleon_weights} shows $\tilde{p}^{(N,1)}$ 
for the nucleon operator (cf. the left-hand panel of 
Fig.~\ref{fig:baryon_elementals} for $\tilde{p}^N$). Clearly the probability 
mass function becomes flatter when marginalised over an increasing number of 
indices. In $\tilde{p}^{N,2}$ all the original structure is lost and the 
  measured optimal probabilities are independent of the index in the localised basis.
Fig.~\ref{fig:exp_N} shows the expected occupancies $\rho^N_k$ of the index 
sets and the cost coefficients $\zeta_k$ of the temporaries as a function of the sample size in the computation of the nucleon-nucleon correlator. Note 
that $\rho^N_1$ reaches its plateau just below $1$ as the 
contraction of the colour components with the epsilon tensor in the construction of the nucleon operator causes the probabilities for any pair of identical 
indices to vanish. From the right-hand side of 
Fig.~\ref{fig:exp_N} we see the cost coefficient of the second temporary plateaus much earlier compared to the other two. Beyond a sample size of $n_s \approx 10^5$ local distillation does not grant any substantial efficiency gain over standard 
distillation in this scenario. 

\subsection{Dependence on the number of constituent quarks and the physical volume}
\label{subsec:tetraquark}

The previous section demonstrated that within the sequence of temporaries, the 
central ones will be the most expensive to 
compute. Explicitly, if the number of constituent quarks $d$ is odd, the $[(d+1)/2]$th temporary will be the 
most expensive. If $d$ is even, the $d/2$th and $(d/2+1)$st temporaries will 
be the most expensive. For example, in the case of a two-point tetraquark correlator, this 
corresponds to the second and third temporaries. Based on our baryon results, we attempt to make a crude prediction
for the cost reduction of a tetraquark calculation using the localised distillation basis and our sampling method.

\begin{figure*}[!thb]
	\centering
	\begin{subfigure}[l]{0.45\textwidth}
		\resizebox{\textwidth}{!}{
			\input{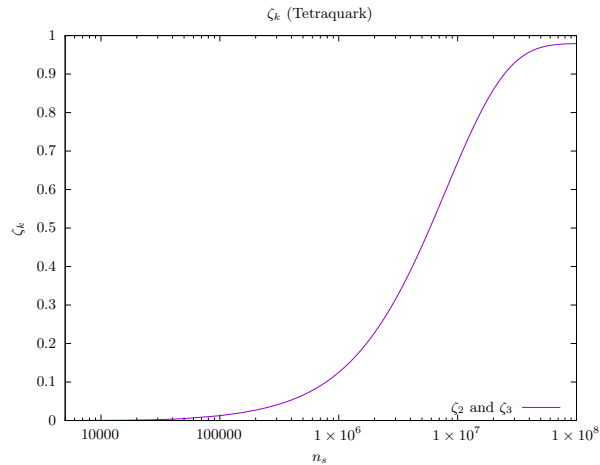}
		}
	\end{subfigure}
	\caption{Expected cost coefficients of the second and third temporary in the computation of the connected tetraquark 2-point correlator}
	\label{fig:exp_N_tetra}
\end{figure*}

Tetraquark operators are typically constructed by contracting a diquark and an 
anti-diquark to a color singlet. The two possible colorless contractions are
$\underline{\bar{3}} \otimes \underline{3}$ and $\underline{6} \otimes \underline{\bar{6}}$. Here we use operators made of (anti-)diquarks in the $\underline{3}$ ($\underline{\bar{3}}$) representation constructed from locally-distilled quark fields with no derivatives between them. 
The quark and anti-quark color indices of this construction will be pairwise 
totally anti-symmetric while their spatial components are symmetric. 
In local distillation, this translates to a tensor of probabilities $p^T_{abcd}$ that is symmetric under distillation-index permutations and zero for any pair 
of identical quark or anti-quark indices. 
Marginalising over one of the indices we expect to obtain a three-index probability tensor similar 
to the nucleon so we use this to estimate the scaling of our algorithm for 
tetraquarks.
To estimate the cost factors of the 2nd and 3rd temporary in the computation of the Wick contractions for the connected tetraquark 2-point correlator, we need the marginalised probabilities $\tilde{p}^{(T,1)} \approx \tilde{p}^{(N,0)}$ and $\tilde{p}^{(T,2)} \approx \tilde{p}^{(N,1)}$. The cost factors $\zeta_2 = \zeta_3$ 
can then be calculated as described in the previous section. 
Figure~\ref{fig:exp_N_tetra} shows the expected scaling of this hypothetical 
computation with the sample size. The final question is which sample size 
would be required for a compact tetraquark operator to achieve a sampling noise 
smaller than the gauge noise.
The answer is not trivial and needs a dedicated investigation, but it is reasonable to assume the necessary sample size scales like $M^{d-1}$, where 
$M$ is the number of sources that significantly overlap with a given 
coarse grid site. This number will be larger when derivatives are applied to 
the sources. In the case of the nucleon we found a calculation with 
$n_s = 5 \times 10^4$ still results in a good signal-to-noise ratio. 
We can approximate $M \approx \sqrt{n_s/\nD} \approx 16$ (we divide by $\nD$ because the location of the first quark source is arbitrary). Since 
both the nucleon and the compact tetraquark operators are ultra-local we 
expect their value of $M$ to be similar. Our rough estimate for the 
minimum sample size needed to compute a tetraquark correlation function is 
then $n_s^T = \nD \times M^3 \approx 800,000$. We see from 
figure~\ref{fig:exp_N_tetra} we may expect a speed-up by a factor of $\approx10$ with respect to full distillation for
this calculation.

Apart from the number of constituent quarks the physical volume, which correlates with the number of grid points, is expected to affect the sparsity of the tensors and thereby the efficiency of the sampling method.
The sparse basis is constructed by first defining a coarse grid of source 
points. If the three-dimensional volume of the lattice is increased and the
physical separation between sites on the basis grid is fixed, this leads 
directly to a linear increase in the number of sites on the grid. To support 
this increase in grid points, the corresponding dimension of the distillation
space must grow in the same proportion. This matches the observed dependence
of the space needed to maintain the same level of smearing in a larger volume. 
It is also commonly noted that a fixed size of distillation space leads to 
a constant level of smearing as the lattice spacing is reduced and this 
corresponds to maintaining a fixed distance in physical units between basis 
grid points. 
Increasing the physical volume and hence the number of coarse grid sites would decrease
the fraction of overlapping sources corresponding to tensor elements with large magnitudes
and is therefore expected to give a greater advantage to local over standard distillation.

\section{Conclusion}
	\label{sec:conclusion}
In this paper, a method to construct a new basis for distillation space, the 
vector space spanned by the lowest eigenvectors of the gauge-covariant 
three-dimensional Laplacian on each time-slice, is introduced. The new basis vectors are  more localised. Methods to 
exploit this locality to accelerate evaluation of the correlation functions 
needed in hadron spectroscopy calculations are then described and tested. 

The eigenvectors of the gauged Laplacian themselves 
form an orthonormal basis but there is no natural locality in these vectors by 
construction. The eigenvectors of the free Laplacian, for example, are the 
Fourier modes and so not localised at all. Something analogous to a 
gauge-covariant wavelet basis combining smoothness and locality would be 
useful and such a scheme was presented and tested in Section~\ref{sec:basis}. 

The method works by applying the distillation space projection operator to
point sources well-separated on a large grid over the time-slice. Empirically,
the resulting vectors stay localised to the sources due to both confinement and
the truncation to a few low-eigenvalue modes in the distillation operator
itself. These vectors are no longer orthonormal, however, and a simple
reorthogonalisation scheme such as Gram-Schmidt destroys locality.
Orthonormality is introduced by defining a flow equation in the space of square
matrices whose stable fixed-point is the subset of unitary matrices. In tests,
it is seen that flowing to the fixed subset makes changes to the
matrix, and subsequently the new basis vectors, which are small enough to 
preserve locality. For 
hadron operators in the new basis, which are colorless tensors made from the vector
components, locality manifests as approximate sparseness where a small subset of entries 
have much larger magnitude compared to the bulk. Correlation functions are unchanged since all that is being
exploited is a basis transformation. The first tests of these ideas in Section~\ref{sec:basis} demonstrated the flow method works well. The sparseness of the tensors suggests a stochastic
importance sampling method might converge quickly and reduce the numerical cost
of tensor contractions.

In Section~\ref{sec:sum}, stochastic representations of sums were introduced, based on
the Hansen-Hurwitz estimator. This uses importance sampling to translate 
the numerical problem of evaluating a large sum, which in this case is 
over distillation space basis vectors, into a more efficient 
Monte Carlo method with minimal variance. The variance reduction is enhanced by
ensuring a small subset of terms contributes the bulk of the signal and this
is enabled in the new basis which generates sparse tensors by exploiting 
locality. 

The expense of using a distillation space of size $\nD$ in a direct computation 
of correlation functions grows rapidly as the number of constituent quark 
fields increases. For meson correlation functions, the cost grows in 
proportion to $\nD^3$ while for baryons, this increases to $\nD^4$. The 
price tag of distillation has restricted calculations of baryons and 
tetraquarks to small vector spaces, which are often far from the best choice 
set by the physics.
The natural way to perform the direct evaluation is to store 
temporary tensors after summing a subset of indices, which
for a simple source-sink construction involving $d$ constituent quark fields reduces the 
scaling from the naive $\nD^{2d}$ to the much more tractable $\nD^{d+1}$ quoted above.
We constructed a Hansen-Hurwitz estimator which preserves this scaling behaviour in the large sample-size limit.
Some first tests of this estimator applied to baryon correlation functions were presented in Section~\ref{sec:application} and showed that good results can be achieved by summing a small fraction of terms contributing to the correlation function.

The bottleneck of our algorithm, in the case of baryons, is the construction of the second temporary tensor, which quickly approaches the $\mathcal{O}(\nD^{4})$ limit as the sample size is increased. A detailed discussion of this effect was presented in section~\ref{sec:discussion}. Some crude extrapolations of our baryon results suggest that this saturation of the limit may become less of a concern as larger distillation spaces and operators with more constituent quarks are used. The example of a hypothetical tetraquark calculation was discussed in section~\ref{subsec:tetraquark}.

The index sampling presented in this article is the most straight-forward implementation of importance sampling. We leave it to future work to investigate more efficient index-sampling schemes that might better exploit the approximate sparsity of the distillation-space operators, for example by incorporating known symmetries of the tensors or by making use of stratified sampling techniques. The effect of phase cancellations on the variance and whether it can be mitigated needs to be studied in more detail. Different kinds of coarse grid embeddings may be tested.
We did not consider sparsity in disconnected insertions, but it is a topic that could be investigated in the future.
We also did not report on tests with non-zero momentum operators as the basis for our application in section~\ref{sec:application} did not require such constructions. In the future, further studies will be required.

While Section~\ref{sec:application} shows promising first test results, there is more work needed
to build a robust algorithm and optimise the necessary work-flows for physics
calculations. The stochastic approach has the promise of better scaling at
larger distillation spaces which would arise as larger physical volumes 
are used, as well as better performance as more constituent quarks are added, 
which has severely limited tetraquark calculations using exact evaluations in 
distillation to date.

\FloatBarrier
\acknowledgments
{
	We thank our colleagues within the Hadron Spectrum Collaboration
	(www.hadspec.org), and Frank Winter and Antoni Woss for
        discussions at an early stage of this work.
	RGE acknowledges support from the U.S.~Department of Energy contract
	DE-AC05-06OR23177, under which Jefferson Science Associates, LLC,
	manages and operates Jefferson Lab.
NL acknowledges support by the EU H2020 research and innovation programme under the Staff Exchange grant agreement No-101086085-ASYMMETRY, as well as by the Spanish Ministerio de Ciencia e Innovacion project PID2020-113644GB-I00  and by Generalitat Valenciana through the grant PROMETEO/2019/083.
This work was supported
by the STRONG-2020 project, funded by the European
Community Horizon 2020 research and innovation programme under grant agreement 824093.
	The software codes {\tt Chroma}~\cite{Edwards:2004sx}, {\tt QUDA}~\cite{Clark:2009wm,Babich:2010mu}, {\tt QPhiX}~\cite{Joo:2013lwm}, {\tt QOPQDP}~\cite{Osborn:2010mb,Babich:2010qb},  and {\tt Redstar}~\cite{Chen:2023zyy} were used.
	This work also used clusters at Jefferson Laboratory under the USQCD Initiative and the LQCD ARRA project, and the authors acknowledge support from the U.S. Department of Energy, Office of Science, Office of Advanced Scientific Computing Research and Office of Nuclear Physics, Scientific Discovery through Advanced Computing (SciDAC) program, and the U.S. Department of Energy Exascale Computing Project.
	This research was supported in part under an ALCC award, and used resources of the Oak Ridge Leadership Computing Facility at the Oak Ridge National Laboratory, which is supported by the Office of Science of the U.S. Department of Energy under Contract No. DE-AC05-00OR22725. This research is also part of the Blue Waters sustained-petascale computing project, which is supported by the National Science Foundation (awards OCI-0725070 and ACI-1238993) and the state of Illinois. Blue Waters is a joint effort of the University of Illinois at Urbana-Champaign and its National Center for Supercomputing Applications. This work is also part of the PRAC “Lattice QCD on Blue Waters”. This research used resources of the National Energy Research Scientific Computing Center (NERSC), a DOE Office of Science User Facility supported by the Office of Science of the U.S. Department of Energy under Contract No. DEAC02-05CH11231. The authors acknowledge the Texas Advanced Computing Center (TACC) at The University of Texas at Austin for providing HPC resources that have contributed to the research results reported within this paper.
	Gauge configurations were generated using resources awarded from the U.S. Department of Energy INCITE program at Oak Ridge National Lab, NERSC, the NSF Teragrid at the Texas Advanced Computer Center and the Pittsburgh Supercomputer Center, as well as at Jefferson Lab.
}

\pagebreak


\bibliographystyle{apsrev4-2}
\bibliography{biblio}

\begin{thebibliography}{29}%
\makeatletter
\providecommand \@ifxundefined [1]{%
 \@ifx{#1\undefined}
}%
\providecommand \@ifnum [1]{%
 \ifnum #1\expandafter \@firstoftwo
 \else \expandafter \@secondoftwo
 \fi
}%
\providecommand \@ifx [1]{%
 \ifx #1\expandafter \@firstoftwo
 \else \expandafter \@secondoftwo
 \fi
}%
\providecommand \natexlab [1]{#1}%
\providecommand \enquote  [1]{``#1''}%
\providecommand \bibnamefont  [1]{#1}%
\providecommand \bibfnamefont [1]{#1}%
\providecommand \citenamefont [1]{#1}%
\providecommand \href@noop [0]{\@secondoftwo}%
\providecommand \href [0]{\begingroup \@sanitize@url \@href}%
\providecommand \@href[1]{\@@startlink{#1}\@@href}%
\providecommand \@@href[1]{\endgroup#1\@@endlink}%
\providecommand \@sanitize@url [0]{\catcode `\\12\catcode `\$12\catcode
  `\&12\catcode `\#12\catcode `\^12\catcode `\_12\catcode `\%12\relax}%
\providecommand \@@startlink[1]{}%
\providecommand \@@endlink[0]{}%
\providecommand \url  [0]{\begingroup\@sanitize@url \@url }%
\providecommand \@url [1]{\endgroup\@href {#1}{\urlprefix }}%
\providecommand \urlprefix  [0]{URL }%
\providecommand \Eprint [0]{\href }%
\providecommand \doibase [0]{https://doi.org/}%
\providecommand \selectlanguage [0]{\@gobble}%
\providecommand \bibinfo  [0]{\@secondoftwo}%
\providecommand \bibfield  [0]{\@secondoftwo}%
\providecommand \translation [1]{[#1]}%
\providecommand \BibitemOpen [0]{}%
\providecommand \bibitemStop [0]{}%
\providecommand \bibitemNoStop [0]{.\EOS\space}%
\providecommand \EOS [0]{\spacefactor3000\relax}%
\providecommand \BibitemShut  [1]{\csname bibitem#1\endcsname}%
\let\auto@bib@innerbib\@empty
\bibitem [{\citenamefont {Peardon}\ \emph {et~al.}(2009)\citenamefont
  {Peardon}, \citenamefont {Bulava}, \citenamefont {Foley}, \citenamefont
  {Morningstar}, \citenamefont {Dudek}, \citenamefont {Edwards}, \citenamefont
  {Joo}, \citenamefont {Lin}, \citenamefont {Richards},\ and\ \citenamefont
  {Juge}}]{Peardon:2009gh}%
  \BibitemOpen
  \bibfield  {author} {\bibinfo {author} {\bibfnamefont {M.}~\bibnamefont
  {Peardon}}, \bibinfo {author} {\bibfnamefont {J.}~\bibnamefont {Bulava}},
  \bibinfo {author} {\bibfnamefont {J.}~\bibnamefont {Foley}}, \bibinfo
  {author} {\bibfnamefont {C.}~\bibnamefont {Morningstar}}, \bibinfo {author}
  {\bibfnamefont {J.}~\bibnamefont {Dudek}}, \bibinfo {author} {\bibfnamefont
  {R.~G.}\ \bibnamefont {Edwards}}, \bibinfo {author} {\bibfnamefont
  {B.}~\bibnamefont {Joo}}, \bibinfo {author} {\bibfnamefont {H.-W.}\
  \bibnamefont {Lin}}, \bibinfo {author} {\bibfnamefont {D.~G.}\ \bibnamefont
  {Richards}},\ and\ \bibinfo {author} {\bibfnamefont {K.~J.}\ \bibnamefont
  {Juge}} (\bibinfo {collaboration} {Hadron Spectrum}),\ }\href
  {https://doi.org/10.1103/PhysRevD.80.054506} {\bibfield  {journal} {\bibinfo
  {journal} {Phys. Rev.}\ }\textbf {\bibinfo {volume} {D80}},\ \bibinfo {pages}
  {054506} (\bibinfo {year} {2009})},\ \Eprint
  {https://arxiv.org/abs/0905.2160} {arXiv:0905.2160 [hep-lat]} \BibitemShut
  {NoStop}%
\bibitem [{\citenamefont {Dudek}\ \emph {et~al.}(2009)\citenamefont {Dudek},
  \citenamefont {Edwards}, \citenamefont {Peardon}, \citenamefont {Richards},\
  and\ \citenamefont {Thomas}}]{Dudek:2009qf}%
  \BibitemOpen
  \bibfield  {author} {\bibinfo {author} {\bibfnamefont {J.~J.}\ \bibnamefont
  {Dudek}}, \bibinfo {author} {\bibfnamefont {R.~G.}\ \bibnamefont {Edwards}},
  \bibinfo {author} {\bibfnamefont {M.~J.}\ \bibnamefont {Peardon}}, \bibinfo
  {author} {\bibfnamefont {D.~G.}\ \bibnamefont {Richards}},\ and\ \bibinfo
  {author} {\bibfnamefont {C.~E.}\ \bibnamefont {Thomas}},\ }\href
  {https://doi.org/10.1103/PhysRevLett.103.262001} {\bibfield  {journal}
  {\bibinfo  {journal} {Phys. Rev. Lett.}\ }\textbf {\bibinfo {volume} {103}},\
  \bibinfo {pages} {262001} (\bibinfo {year} {2009})},\ \Eprint
  {https://arxiv.org/abs/0909.0200} {arXiv:0909.0200 [hep-ph]} \BibitemShut
  {NoStop}%
\bibitem [{\citenamefont {Dudek}\ \emph {et~al.}(2011)\citenamefont {Dudek},
  \citenamefont {Edwards}, \citenamefont {Joo}, \citenamefont {Peardon},
  \citenamefont {Richards},\ and\ \citenamefont {Thomas}}]{Dudek:2011tt}%
  \BibitemOpen
  \bibfield  {author} {\bibinfo {author} {\bibfnamefont {J.~J.}\ \bibnamefont
  {Dudek}}, \bibinfo {author} {\bibfnamefont {R.~G.}\ \bibnamefont {Edwards}},
  \bibinfo {author} {\bibfnamefont {B.}~\bibnamefont {Joo}}, \bibinfo {author}
  {\bibfnamefont {M.~J.}\ \bibnamefont {Peardon}}, \bibinfo {author}
  {\bibfnamefont {D.~G.}\ \bibnamefont {Richards}},\ and\ \bibinfo {author}
  {\bibfnamefont {C.~E.}\ \bibnamefont {Thomas}},\ }\href
  {https://doi.org/10.1103/PhysRevD.83.111502} {\bibfield  {journal} {\bibinfo
  {journal} {Phys. Rev. D}\ }\textbf {\bibinfo {volume} {83}},\ \bibinfo
  {pages} {111502} (\bibinfo {year} {2011})},\ \Eprint
  {https://arxiv.org/abs/1102.4299} {arXiv:1102.4299 [hep-lat]} \BibitemShut
  {NoStop}%
\bibitem [{\citenamefont {Liu}\ \emph {et~al.}(2012)\citenamefont {Liu},
  \citenamefont {Moir}, \citenamefont {Peardon}, \citenamefont {Ryan},
  \citenamefont {Thomas}, \citenamefont {Vilaseca}, \citenamefont {Dudek},
  \citenamefont {Edwards}, \citenamefont {Joo},\ and\ \citenamefont
  {Richards}}]{HadronSpectrum:2012gic}%
  \BibitemOpen
  \bibfield  {author} {\bibinfo {author} {\bibfnamefont {L.}~\bibnamefont
  {Liu}}, \bibinfo {author} {\bibfnamefont {G.}~\bibnamefont {Moir}}, \bibinfo
  {author} {\bibfnamefont {M.}~\bibnamefont {Peardon}}, \bibinfo {author}
  {\bibfnamefont {S.~M.}\ \bibnamefont {Ryan}}, \bibinfo {author}
  {\bibfnamefont {C.~E.}\ \bibnamefont {Thomas}}, \bibinfo {author}
  {\bibfnamefont {P.}~\bibnamefont {Vilaseca}}, \bibinfo {author}
  {\bibfnamefont {J.~J.}\ \bibnamefont {Dudek}}, \bibinfo {author}
  {\bibfnamefont {R.~G.}\ \bibnamefont {Edwards}}, \bibinfo {author}
  {\bibfnamefont {B.}~\bibnamefont {Joo}},\ and\ \bibinfo {author}
  {\bibfnamefont {D.~G.}\ \bibnamefont {Richards}} (\bibinfo {collaboration}
  {Hadron Spectrum}),\ }\href {https://doi.org/10.1007/JHEP07(2012)126}
  {\bibfield  {journal} {\bibinfo  {journal} {JHEP}\ }\textbf {\bibinfo
  {volume} {07}},\ \bibinfo {pages} {126}},\ \Eprint
  {https://arxiv.org/abs/1204.5425} {arXiv:1204.5425 [hep-ph]} \BibitemShut
  {NoStop}%
\bibitem [{\citenamefont {Lang}\ and\ \citenamefont
  {Wilson}(2022)}]{Lang:2022elg}%
  \BibitemOpen
  \bibfield  {author} {\bibinfo {author} {\bibfnamefont {N.}~\bibnamefont
  {Lang}}\ and\ \bibinfo {author} {\bibfnamefont {D.~J.}\ \bibnamefont
  {Wilson}} (\bibinfo {collaboration} {Hadron Spectrum}),\ }\href
  {https://doi.org/10.1103/PhysRevLett.129.252001} {\bibfield  {journal}
  {\bibinfo  {journal} {Phys. Rev. Lett.}\ }\textbf {\bibinfo {volume} {129}},\
  \bibinfo {pages} {252001} (\bibinfo {year} {2022})},\ \Eprint
  {https://arxiv.org/abs/2205.05026} {arXiv:2205.05026 [hep-ph]} \BibitemShut
  {NoStop}%
\bibitem [{\citenamefont {Wilson}\ \emph {et~al.}(2024)\citenamefont {Wilson},
  \citenamefont {Thomas}, \citenamefont {Dudek},\ and\ \citenamefont
  {Edwards}}]{Wilson:2023hzu}%
  \BibitemOpen
  \bibfield  {author} {\bibinfo {author} {\bibfnamefont {D.~J.}\ \bibnamefont
  {Wilson}}, \bibinfo {author} {\bibfnamefont {C.~E.}\ \bibnamefont {Thomas}},
  \bibinfo {author} {\bibfnamefont {J.~J.}\ \bibnamefont {Dudek}},\ and\
  \bibinfo {author} {\bibfnamefont {R.~G.}\ \bibnamefont {Edwards}} (\bibinfo
  {collaboration} {Hadron Spectrum}),\ }\href
  {https://doi.org/10.1103/PhysRevLett.132.241901} {\bibfield  {journal}
  {\bibinfo  {journal} {Phys. Rev. Lett.}\ }\textbf {\bibinfo {volume} {132}},\
  \bibinfo {pages} {241901} (\bibinfo {year} {2024})},\ \Eprint
  {https://arxiv.org/abs/2309.14070} {arXiv:2309.14070 [hep-lat]} \BibitemShut
  {NoStop}%
\bibitem [{\citenamefont {Yan}\ \emph {et~al.}(2024)\citenamefont {Yan},
  \citenamefont {Garofalo}, \citenamefont {Mai}, \citenamefont {Mei\ss{}ner},\
  and\ \citenamefont {Urbach}}]{Yan:2024gwp}%
  \BibitemOpen
  \bibfield  {author} {\bibinfo {author} {\bibfnamefont {H.}~\bibnamefont
  {Yan}}, \bibinfo {author} {\bibfnamefont {M.}~\bibnamefont {Garofalo}},
  \bibinfo {author} {\bibfnamefont {M.}~\bibnamefont {Mai}}, \bibinfo {author}
  {\bibfnamefont {U.-G.}\ \bibnamefont {Mei\ss{}ner}},\ and\ \bibinfo {author}
  {\bibfnamefont {C.}~\bibnamefont {Urbach}},\ }\href@noop {} {\  (\bibinfo
  {year} {2024})},\ \Eprint {https://arxiv.org/abs/2407.16659}
  {arXiv:2407.16659 [hep-lat]} \BibitemShut {NoStop}%
\bibitem [{\citenamefont {Li}\ \emph {et~al.}(2024)\citenamefont {Li},
  \citenamefont {Shi}, \citenamefont {Chen}, \citenamefont {Gong},
  \citenamefont {Liang}, \citenamefont {Liu},\ and\ \citenamefont
  {Sun}}]{Li:2024pfg}%
  \BibitemOpen
  \bibfield  {author} {\bibinfo {author} {\bibfnamefont {H.}~\bibnamefont
  {Li}}, \bibinfo {author} {\bibfnamefont {C.}~\bibnamefont {Shi}}, \bibinfo
  {author} {\bibfnamefont {Y.}~\bibnamefont {Chen}}, \bibinfo {author}
  {\bibfnamefont {M.}~\bibnamefont {Gong}}, \bibinfo {author} {\bibfnamefont
  {J.}~\bibnamefont {Liang}}, \bibinfo {author} {\bibfnamefont
  {Z.}~\bibnamefont {Liu}},\ and\ \bibinfo {author} {\bibfnamefont
  {W.}~\bibnamefont {Sun}},\ }\href@noop {} {\  (\bibinfo {year} {2024})},\
  \Eprint {https://arxiv.org/abs/2402.14541} {arXiv:2402.14541 [hep-lat]}
  \BibitemShut {NoStop}%
\bibitem [{\citenamefont {Boyle}\ \emph {et~al.}()\citenamefont {Boyle},
  \citenamefont {Erben}, \citenamefont {G\"ulpers}, \citenamefont {Hansen},
  \citenamefont {Joswig}, \citenamefont {Lachini}, \citenamefont {Marshall},\
  and\ \citenamefont {Portelli}}]{Boyle:2024hvv}%
  \BibitemOpen
  \bibfield  {author} {\bibinfo {author} {\bibfnamefont {P.}~\bibnamefont
  {Boyle}}, \bibinfo {author} {\bibfnamefont {F.}~\bibnamefont {Erben}},
  \bibinfo {author} {\bibfnamefont {V.}~\bibnamefont {G\"ulpers}}, \bibinfo
  {author} {\bibfnamefont {M.~T.}\ \bibnamefont {Hansen}}, \bibinfo {author}
  {\bibfnamefont {F.}~\bibnamefont {Joswig}}, \bibinfo {author} {\bibfnamefont
  {N.~P.}\ \bibnamefont {Lachini}}, \bibinfo {author} {\bibfnamefont
  {M.}~\bibnamefont {Marshall}},\ and\ \bibinfo {author} {\bibfnamefont
  {A.}~\bibnamefont {Portelli}},\ }\href@noop {} {\ }\Eprint
  {https://arxiv.org/abs/2406.19194} {arXiv:2406.19194 [hep-lat]} \BibitemShut
  {NoStop}%
\bibitem [{\citenamefont {Knechtli}\ \emph {et~al.}(2022)\citenamefont
  {Knechtli}, \citenamefont {Korzec}, \citenamefont {Peardon},\ and\
  \citenamefont {Urrea-Ni\~no}}]{Knechtli:2022bji}%
  \BibitemOpen
  \bibfield  {author} {\bibinfo {author} {\bibfnamefont {F.}~\bibnamefont
  {Knechtli}}, \bibinfo {author} {\bibfnamefont {T.}~\bibnamefont {Korzec}},
  \bibinfo {author} {\bibfnamefont {M.}~\bibnamefont {Peardon}},\ and\ \bibinfo
  {author} {\bibfnamefont {J.~A.}\ \bibnamefont {Urrea-Ni\~no}},\ }\href
  {https://doi.org/10.1103/PhysRevD.106.034501} {\bibfield  {journal} {\bibinfo
   {journal} {Phys. Rev. D}\ }\textbf {\bibinfo {volume} {106}},\ \bibinfo
  {pages} {034501} (\bibinfo {year} {2022})},\ \Eprint
  {https://arxiv.org/abs/2205.11564} {arXiv:2205.11564 [hep-lat]} \BibitemShut
  {NoStop}%
\bibitem [{\citenamefont {Luscher}(1991)}]{Luscher:1990ux}%
  \BibitemOpen
  \bibfield  {author} {\bibinfo {author} {\bibfnamefont {M.}~\bibnamefont
  {Luscher}},\ }\href {https://doi.org/10.1016/0550-3213(91)90366-6} {\bibfield
   {journal} {\bibinfo  {journal} {Nucl. Phys. B}\ }\textbf {\bibinfo {volume}
  {354}},\ \bibinfo {pages} {531} (\bibinfo {year} {1991})}\BibitemShut
  {NoStop}%
\bibitem [{\citenamefont {Luscher}\ and\ \citenamefont
  {Wolff}(1990)}]{Luscher:1990ck}%
  \BibitemOpen
  \bibfield  {author} {\bibinfo {author} {\bibfnamefont {M.}~\bibnamefont
  {Luscher}}\ and\ \bibinfo {author} {\bibfnamefont {U.}~\bibnamefont
  {Wolff}},\ }\href {https://doi.org/10.1016/0550-3213(90)90540-T} {\bibfield
  {journal} {\bibinfo  {journal} {Nucl. Phys. B}\ }\textbf {\bibinfo {volume}
  {339}},\ \bibinfo {pages} {222} (\bibinfo {year} {1990})}\BibitemShut
  {NoStop}%
\bibitem [{\citenamefont {Briceno}\ \emph {et~al.}(2018)\citenamefont
  {Briceno}, \citenamefont {Dudek},\ and\ \citenamefont
  {Young}}]{Briceno:2017max}%
  \BibitemOpen
  \bibfield  {author} {\bibinfo {author} {\bibfnamefont {R.~A.}\ \bibnamefont
  {Briceno}}, \bibinfo {author} {\bibfnamefont {J.~J.}\ \bibnamefont {Dudek}},\
  and\ \bibinfo {author} {\bibfnamefont {R.~D.}\ \bibnamefont {Young}},\ }\href
  {https://doi.org/10.1103/RevModPhys.90.025001} {\bibfield  {journal}
  {\bibinfo  {journal} {Rev. Mod. Phys.}\ }\textbf {\bibinfo {volume} {90}},\
  \bibinfo {pages} {025001} (\bibinfo {year} {2018})},\ \Eprint
  {https://arxiv.org/abs/1706.06223} {arXiv:1706.06223 [hep-lat]} \BibitemShut
  {NoStop}%
\bibitem [{\citenamefont {Basak}\ \emph {et~al.}(2005)\citenamefont {Basak},
  \citenamefont {Edwards}, \citenamefont {Fleming}, \citenamefont {Heller},
  \citenamefont {Morningstar}, \citenamefont {Richards}, \citenamefont {Sato},\
  and\ \citenamefont {Wallace}}]{Basak:2005aq}%
  \BibitemOpen
  \bibfield  {author} {\bibinfo {author} {\bibfnamefont {S.}~\bibnamefont
  {Basak}}, \bibinfo {author} {\bibfnamefont {R.~G.}\ \bibnamefont {Edwards}},
  \bibinfo {author} {\bibfnamefont {G.~T.}\ \bibnamefont {Fleming}}, \bibinfo
  {author} {\bibfnamefont {U.~M.}\ \bibnamefont {Heller}}, \bibinfo {author}
  {\bibfnamefont {C.}~\bibnamefont {Morningstar}}, \bibinfo {author}
  {\bibfnamefont {D.}~\bibnamefont {Richards}}, \bibinfo {author}
  {\bibfnamefont {I.}~\bibnamefont {Sato}},\ and\ \bibinfo {author}
  {\bibfnamefont {S.}~\bibnamefont {Wallace}},\ }\href
  {https://doi.org/10.1103/PhysRevD.72.094506} {\bibfield  {journal} {\bibinfo
  {journal} {Phys. Rev. D}\ }\textbf {\bibinfo {volume} {72}},\ \bibinfo
  {pages} {094506} (\bibinfo {year} {2005})},\ \Eprint
  {https://arxiv.org/abs/hep-lat/0506029} {arXiv:hep-lat/0506029} \BibitemShut
  {NoStop}%
\bibitem [{\citenamefont {Edwards}\ \emph {et~al.}(2011)\citenamefont
  {Edwards}, \citenamefont {Dudek}, \citenamefont {Richards},\ and\
  \citenamefont {Wallace}}]{Edwards:2011jj}%
  \BibitemOpen
  \bibfield  {author} {\bibinfo {author} {\bibfnamefont {R.~G.}\ \bibnamefont
  {Edwards}}, \bibinfo {author} {\bibfnamefont {J.~J.}\ \bibnamefont {Dudek}},
  \bibinfo {author} {\bibfnamefont {D.~G.}\ \bibnamefont {Richards}},\ and\
  \bibinfo {author} {\bibfnamefont {S.~J.}\ \bibnamefont {Wallace}},\ }\href
  {https://doi.org/10.1103/PhysRevD.84.074508} {\bibfield  {journal} {\bibinfo
  {journal} {Phys. Rev. D}\ }\textbf {\bibinfo {volume} {84}},\ \bibinfo
  {pages} {074508} (\bibinfo {year} {2011})},\ \Eprint
  {https://arxiv.org/abs/1104.5152} {arXiv:1104.5152 [hep-ph]} \BibitemShut
  {NoStop}%
\bibitem [{\citenamefont {Green}\ \emph {et~al.}(2021)\citenamefont {Green},
  \citenamefont {Hanlon}, \citenamefont {Junnarkar},\ and\ \citenamefont
  {Wittig}}]{Green:2021qol}%
  \BibitemOpen
  \bibfield  {author} {\bibinfo {author} {\bibfnamefont {J.~R.}\ \bibnamefont
  {Green}}, \bibinfo {author} {\bibfnamefont {A.~D.}\ \bibnamefont {Hanlon}},
  \bibinfo {author} {\bibfnamefont {P.~M.}\ \bibnamefont {Junnarkar}},\ and\
  \bibinfo {author} {\bibfnamefont {H.}~\bibnamefont {Wittig}},\ }\href
  {https://doi.org/10.1103/PhysRevLett.127.242003} {\bibfield  {journal}
  {\bibinfo  {journal} {Phys. Rev. Lett.}\ }\textbf {\bibinfo {volume} {127}},\
  \bibinfo {pages} {242003} (\bibinfo {year} {2021})},\ \Eprint
  {https://arxiv.org/abs/2103.01054} {arXiv:2103.01054 [hep-lat]} \BibitemShut
  {NoStop}%
\bibitem [{\citenamefont {Cheung}\ \emph {et~al.}(2017)\citenamefont {Cheung},
  \citenamefont {Thomas}, \citenamefont {Dudek},\ and\ \citenamefont
  {Edwards}}]{Cheung:2017tnt}%
  \BibitemOpen
  \bibfield  {author} {\bibinfo {author} {\bibfnamefont {G.~K.~C.}\
  \bibnamefont {Cheung}}, \bibinfo {author} {\bibfnamefont {C.~E.}\
  \bibnamefont {Thomas}}, \bibinfo {author} {\bibfnamefont {J.~J.}\
  \bibnamefont {Dudek}},\ and\ \bibinfo {author} {\bibfnamefont {R.~G.}\
  \bibnamefont {Edwards}} (\bibinfo {collaboration} {Hadron Spectrum}),\ }\href
  {https://doi.org/10.1007/JHEP11(2017)033} {\bibfield  {journal} {\bibinfo
  {journal} {JHEP}\ }\textbf {\bibinfo {volume} {11}},\ \bibinfo {pages}
  {033}},\ \Eprint {https://arxiv.org/abs/1709.01417} {arXiv:1709.01417
  [hep-lat]} \BibitemShut {NoStop}%
\bibitem [{\citenamefont {Collins}\ \emph {et~al.}(2024)\citenamefont
  {Collins}, \citenamefont {Nefediev}, \citenamefont {Padmanath},\ and\
  \citenamefont {Prelovsek}}]{Collins:2024sfi}%
  \BibitemOpen
  \bibfield  {author} {\bibinfo {author} {\bibfnamefont {S.}~\bibnamefont
  {Collins}}, \bibinfo {author} {\bibfnamefont {A.}~\bibnamefont {Nefediev}},
  \bibinfo {author} {\bibfnamefont {M.}~\bibnamefont {Padmanath}},\ and\
  \bibinfo {author} {\bibfnamefont {S.}~\bibnamefont {Prelovsek}},\ }\href
  {https://doi.org/10.1103/PhysRevD.109.094509} {\bibfield  {journal} {\bibinfo
   {journal} {Phys. Rev. D}\ }\textbf {\bibinfo {volume} {109}},\ \bibinfo
  {pages} {094509} (\bibinfo {year} {2024})},\ \Eprint
  {https://arxiv.org/abs/2402.14715} {arXiv:2402.14715 [hep-lat]} \BibitemShut
  {NoStop}%
\bibitem [{\citenamefont {Morningstar}\ \emph {et~al.}(2011)\citenamefont
  {Morningstar}, \citenamefont {Bulava}, \citenamefont {Foley}, \citenamefont
  {Juge}, \citenamefont {Lenkner}, \citenamefont {Peardon},\ and\ \citenamefont
  {Wong}}]{Morningstar:2011ka}%
  \BibitemOpen
  \bibfield  {author} {\bibinfo {author} {\bibfnamefont {C.}~\bibnamefont
  {Morningstar}}, \bibinfo {author} {\bibfnamefont {J.}~\bibnamefont {Bulava}},
  \bibinfo {author} {\bibfnamefont {J.}~\bibnamefont {Foley}}, \bibinfo
  {author} {\bibfnamefont {K.~J.}\ \bibnamefont {Juge}}, \bibinfo {author}
  {\bibfnamefont {D.}~\bibnamefont {Lenkner}}, \bibinfo {author} {\bibfnamefont
  {M.}~\bibnamefont {Peardon}},\ and\ \bibinfo {author} {\bibfnamefont {C.~H.}\
  \bibnamefont {Wong}},\ }\href {https://doi.org/10.1103/PhysRevD.83.114505}
  {\bibfield  {journal} {\bibinfo  {journal} {Phys. Rev. D}\ }\textbf {\bibinfo
  {volume} {83}},\ \bibinfo {pages} {114505} (\bibinfo {year} {2011})},\
  \Eprint {https://arxiv.org/abs/1104.3870} {arXiv:1104.3870 [hep-lat]}
  \BibitemShut {NoStop}%
\bibitem [{\citenamefont {Foley}\ \emph {et~al.}(2005)\citenamefont {Foley},
  \citenamefont {Jimmy~Juge}, \citenamefont {O'Cais}, \citenamefont {Peardon},
  \citenamefont {Ryan},\ and\ \citenamefont {Skullerud}}]{Foley:2005ac}%
  \BibitemOpen
  \bibfield  {author} {\bibinfo {author} {\bibfnamefont {J.}~\bibnamefont
  {Foley}}, \bibinfo {author} {\bibfnamefont {K.}~\bibnamefont {Jimmy~Juge}},
  \bibinfo {author} {\bibfnamefont {A.}~\bibnamefont {O'Cais}}, \bibinfo
  {author} {\bibfnamefont {M.}~\bibnamefont {Peardon}}, \bibinfo {author}
  {\bibfnamefont {S.~M.}\ \bibnamefont {Ryan}},\ and\ \bibinfo {author}
  {\bibfnamefont {J.-I.}\ \bibnamefont {Skullerud}},\ }\href
  {https://doi.org/10.1016/j.cpc.2005.06.008} {\bibfield  {journal} {\bibinfo
  {journal} {Comput. Phys. Commun.}\ }\textbf {\bibinfo {volume} {172}},\
  \bibinfo {pages} {145} (\bibinfo {year} {2005})},\ \Eprint
  {https://arxiv.org/abs/hep-lat/0505023} {arXiv:hep-lat/0505023} \BibitemShut
  {NoStop}%
\bibitem [{\citenamefont {Hansen}\ and\ \citenamefont
  {Hurwitz}(1943)}]{hansen1943}%
  \BibitemOpen
  \bibfield  {author} {\bibinfo {author} {\bibfnamefont {M.~H.}\ \bibnamefont
  {Hansen}}\ and\ \bibinfo {author} {\bibfnamefont {W.~N.}\ \bibnamefont
  {Hurwitz}},\ }\href {https://doi.org/10.1214/aoms/1177731356} {\bibfield
  {journal} {\bibinfo  {journal} {The Annals of Mathematical Statistics}\
  }\textbf {\bibinfo {volume} {14}},\ \bibinfo {pages} {333 } (\bibinfo {year}
  {1943})}\BibitemShut {NoStop}%
\bibitem [{\citenamefont {Morningstar}\ and\ \citenamefont
  {Peardon}(2004)}]{Morningstar_2004}%
  \BibitemOpen
  \bibfield  {author} {\bibinfo {author} {\bibfnamefont {C.}~\bibnamefont
  {Morningstar}}\ and\ \bibinfo {author} {\bibfnamefont {M.}~\bibnamefont
  {Peardon}},\ }\bibfield  {journal} {\bibinfo  {journal} {Physical Review D}\
  }\textbf {\bibinfo {volume} {69}},\ \href
  {https://doi.org/10.1103/physrevd.69.054501} {10.1103/physrevd.69.054501}
  (\bibinfo {year} {2004})\BibitemShut {NoStop}%
\bibitem [{\citenamefont {Edwards}\ and\ \citenamefont
  {Joo}(2005)}]{Edwards:2004sx}%
  \BibitemOpen
  \bibfield  {author} {\bibinfo {author} {\bibfnamefont {R.~G.}\ \bibnamefont
  {Edwards}}\ and\ \bibinfo {author} {\bibfnamefont {B.}~\bibnamefont {Joo}}
  (\bibinfo {collaboration} {SciDAC, LHPC, UKQCD}),\ }\href
  {https://doi.org/10.1016/j.nuclphysbps.2004.11.254} {\bibfield  {journal}
  {\bibinfo  {journal} {Nucl. Phys. B Proc. Suppl.}\ }\textbf {\bibinfo
  {volume} {140}},\ \bibinfo {pages} {832} (\bibinfo {year} {2005})},\ \Eprint
  {https://arxiv.org/abs/hep-lat/0409003} {arXiv:hep-lat/0409003} \BibitemShut
  {NoStop}%
\bibitem [{\citenamefont {Clark}\ \emph {et~al.}(2010)\citenamefont {Clark},
  \citenamefont {Babich}, \citenamefont {Barros}, \citenamefont {Brower},\ and\
  \citenamefont {Rebbi}}]{Clark:2009wm}%
  \BibitemOpen
  \bibfield  {author} {\bibinfo {author} {\bibfnamefont {M.~A.}\ \bibnamefont
  {Clark}}, \bibinfo {author} {\bibfnamefont {R.}~\bibnamefont {Babich}},
  \bibinfo {author} {\bibfnamefont {K.}~\bibnamefont {Barros}}, \bibinfo
  {author} {\bibfnamefont {R.~C.}\ \bibnamefont {Brower}},\ and\ \bibinfo
  {author} {\bibfnamefont {C.}~\bibnamefont {Rebbi}} (\bibinfo {collaboration}
  {QUDA}),\ }\href {https://doi.org/10.1016/j.cpc.2010.05.002} {\bibfield
  {journal} {\bibinfo  {journal} {Comput. Phys. Commun.}\ }\textbf {\bibinfo
  {volume} {181}},\ \bibinfo {pages} {1517} (\bibinfo {year} {2010})},\ \Eprint
  {https://arxiv.org/abs/0911.3191} {arXiv:0911.3191 [hep-lat]} \BibitemShut
  {NoStop}%
\bibitem [{\citenamefont {Babich}\ \emph
  {et~al.}(2010{\natexlab{a}})\citenamefont {Babich}, \citenamefont {Clark},\
  and\ \citenamefont {Joo}}]{Babich:2010mu}%
  \BibitemOpen
  \bibfield  {author} {\bibinfo {author} {\bibfnamefont {R.}~\bibnamefont
  {Babich}}, \bibinfo {author} {\bibfnamefont {M.~A.}\ \bibnamefont {Clark}},\
  and\ \bibinfo {author} {\bibfnamefont {B.}~\bibnamefont {Joo}},\ }in\
  \href@noop {} {\emph {\bibinfo {booktitle} {{SC 10 (Supercomputing 2010)}}}}\
  (\bibinfo {year} {2010})\ \Eprint {https://arxiv.org/abs/1011.0024}
  {arXiv:1011.0024 [hep-lat]} \BibitemShut {NoStop}%
\bibitem [{\citenamefont {Jo\'o}\ \emph {et~al.}(2013)\citenamefont {Jo\'o},
  \citenamefont {Kalamkar}, \citenamefont {Vaidyanathan}, \citenamefont
  {Smelyanskiy}, \citenamefont {Pamnany}, \citenamefont {Lee}, \citenamefont
  {Dubey},\ and\ \citenamefont {Watson}}]{Joo:2013lwm}%
  \BibitemOpen
  \bibfield  {author} {\bibinfo {author} {\bibfnamefont {B.}~\bibnamefont
  {Jo\'o}}, \bibinfo {author} {\bibfnamefont {D.~D.}\ \bibnamefont {Kalamkar}},
  \bibinfo {author} {\bibfnamefont {K.}~\bibnamefont {Vaidyanathan}}, \bibinfo
  {author} {\bibfnamefont {M.}~\bibnamefont {Smelyanskiy}}, \bibinfo {author}
  {\bibfnamefont {K.}~\bibnamefont {Pamnany}}, \bibinfo {author} {\bibfnamefont
  {V.~W.}\ \bibnamefont {Lee}}, \bibinfo {author} {\bibfnamefont
  {P.}~\bibnamefont {Dubey}},\ and\ \bibinfo {author} {\bibfnamefont
  {W.}~\bibnamefont {Watson}},\ }\href
  {https://doi.org/10.1007/978-3-642-38750-0_4} {\bibfield  {journal} {\bibinfo
   {journal} {Lect. Notes Comput. Sci.}\ }\textbf {\bibinfo {volume} {7905}},\
  \bibinfo {pages} {40} (\bibinfo {year} {2013})}\BibitemShut {NoStop}%
\bibitem [{\citenamefont {Osborn}\ \emph {et~al.}(2010)\citenamefont {Osborn},
  \citenamefont {Babich}, \citenamefont {Brannick}, \citenamefont {Brower},
  \citenamefont {Clark}, \citenamefont {Cohen},\ and\ \citenamefont
  {Rebbi}}]{Osborn:2010mb}%
  \BibitemOpen
  \bibfield  {author} {\bibinfo {author} {\bibfnamefont {J.~C.}\ \bibnamefont
  {Osborn}}, \bibinfo {author} {\bibfnamefont {R.}~\bibnamefont {Babich}},
  \bibinfo {author} {\bibfnamefont {J.}~\bibnamefont {Brannick}}, \bibinfo
  {author} {\bibfnamefont {R.~C.}\ \bibnamefont {Brower}}, \bibinfo {author}
  {\bibfnamefont {M.~A.}\ \bibnamefont {Clark}}, \bibinfo {author}
  {\bibfnamefont {S.~D.}\ \bibnamefont {Cohen}},\ and\ \bibinfo {author}
  {\bibfnamefont {C.}~\bibnamefont {Rebbi}},\ }\href
  {https://doi.org/10.22323/1.105.0037} {\bibfield  {journal} {\bibinfo
  {journal} {PoS}\ }\textbf {\bibinfo {volume} {LATTICE2010}},\ \bibinfo
  {pages} {037} (\bibinfo {year} {2010})},\ \Eprint
  {https://arxiv.org/abs/1011.2775} {arXiv:1011.2775 [hep-lat]} \BibitemShut
  {NoStop}%
\bibitem [{\citenamefont {Babich}\ \emph
  {et~al.}(2010{\natexlab{b}})\citenamefont {Babich}, \citenamefont {Brannick},
  \citenamefont {Brower}, \citenamefont {Clark}, \citenamefont {Manteuffel},
  \citenamefont {McCormick}, \citenamefont {Osborn},\ and\ \citenamefont
  {Rebbi}}]{Babich:2010qb}%
  \BibitemOpen
  \bibfield  {author} {\bibinfo {author} {\bibfnamefont {R.}~\bibnamefont
  {Babich}}, \bibinfo {author} {\bibfnamefont {J.}~\bibnamefont {Brannick}},
  \bibinfo {author} {\bibfnamefont {R.~C.}\ \bibnamefont {Brower}}, \bibinfo
  {author} {\bibfnamefont {M.~A.}\ \bibnamefont {Clark}}, \bibinfo {author}
  {\bibfnamefont {T.~A.}\ \bibnamefont {Manteuffel}}, \bibinfo {author}
  {\bibfnamefont {S.~F.}\ \bibnamefont {McCormick}}, \bibinfo {author}
  {\bibfnamefont {J.~C.}\ \bibnamefont {Osborn}},\ and\ \bibinfo {author}
  {\bibfnamefont {C.}~\bibnamefont {Rebbi}},\ }\href
  {https://doi.org/10.1103/PhysRevLett.105.201602} {\bibfield  {journal}
  {\bibinfo  {journal} {Phys. Rev. Lett.}\ }\textbf {\bibinfo {volume} {105}},\
  \bibinfo {pages} {201602} (\bibinfo {year} {2010}{\natexlab{b}})},\ \Eprint
  {https://arxiv.org/abs/1005.3043} {arXiv:1005.3043 [hep-lat]} \BibitemShut
  {NoStop}%
\bibitem [{\citenamefont {Chen}\ \emph {et~al.}(2023)\citenamefont {Chen},
  \citenamefont {Edwards},\ and\ \citenamefont {Mao}}]{Chen:2023zyy}%
  \BibitemOpen
  \bibfield  {author} {\bibinfo {author} {\bibfnamefont {J.}~\bibnamefont
  {Chen}}, \bibinfo {author} {\bibfnamefont {R.~G.}\ \bibnamefont {Edwards}},\
  and\ \bibinfo {author} {\bibfnamefont {W.}~\bibnamefont {Mao}},\ }in\ \href
  {https://doi.org/10.1145/3592979.3593409} {\emph {\bibinfo {booktitle}
  {{Platform for Advanced Scientific Computing}}}}\ (\bibinfo {year}
  {2023})\BibitemShut {NoStop}%
\end{thebibliography}%


\end{document}